\documentclass[amsmath,amssymb,reprint,physrev,citeautoscript,floatfix,superscriptaddress,noeprint]{revtex4-2}

\usepackage{graphicx}
\usepackage{dcolumn}
\usepackage{bm}
\usepackage[group-separator={,}]{siunitx}
\usepackage[breaklinks,colorlinks,linkcolor={blue}, %
            citecolor={blue},urlcolor={blue}]{hyperref}
\usepackage[version=4]{mhchem}
\setcitestyle{super}
\usepackage{xcolor}

\begin{document}

\title{A map of single-phase high-entropy alloys}
       
\author{Wei Chen}
\affiliation{Institute of Condensed Matter and Nanoscicence (IMCN),
Universit\'{e} catholique de Louvain,
Chemin Etoiles 8, 
Louvain-la-Neuve 1348, Belgium}
\author{Antoine Hilhorst}
\affiliation{UCLouvain, Institute of Mechanics, Materials and Civil Engineering (iMMC), IMAP,
Place Sainte Barbe 2, Louvain-la-Neuve 1348, Belgium}
\author{Georges Bokas}
\affiliation{Institute of Condensed Matter and Nanoscicence (IMCN),
Universit\'{e} catholique de Louvain,
Chemin Etoiles 8, 
Louvain-la-Neuve 1348, Belgium}
\author{St\'{e}phane Gorsse}
\affiliation{CNRS, University of Bordeaux, Bordeaux INP, ICMCB,
UMR 5026,
Pessac 33600, France
}
\author{Pascal J. Jacques}
\affiliation{UCLouvain, Institute of Mechanics, Materials and Civil Engineering (iMMC), IMAP,
Place Sainte Barbe 2, Louvain-la-Neuve 1348, Belgium}
\author{Geoffroy Hautier}
\email[e-mail: ]{geoffroy.t.f.hautier@dartmouth.edu}
\affiliation{Institute of Condensed Matter and Nanoscicence (IMCN),
Universit\'{e} catholique de Louvain,
Chemin Etoiles 8, 
Louvain-la-Neuve 1348, Belgium}
\affiliation{Thayer School of Engineering, Dartmouth College,
Hanover, New Hampshire 03755, USA}

\begin{abstract}
High-entropy alloys have exhibited unusual materials properties. 
The stability of equimolar single-phase solid solution of five or more elements is 
supposedly rare and identifying the existence of such alloys has been challenging 
because of the vast chemical space of possible combinations. 
Herein, based on high-throughput density-functional theory calculations, 
we construct a chemical map of single-phase equimolar high entropy alloys by investigating 
over \num{658000} equimolar quinary alloys through a binary regular solid-solution model. 
We identify {\num{30201}}
potential single-phase equimolar alloys 
(5\% of the possible combinations) forming mainly in body-centered cubic structures. 
We unveil the chemistries that are likely to form high-entropy alloys, 
and identify the complex interplay among mixing enthalpy, intermetallics formation, 
and melting point that drives the formation of these solid solutions. 
We demonstrate the power of our method by predicting the existence of two new high entropy alloys, 
i.e.\ the body-centered cubic \ce{AlCoMnNiV} and the face-centered cubic \ce{CoFeMnNiZn},
which are successfully synthesized.
\end{abstract}

\date{\today}
\maketitle

\section*{Introduction}

The field of metallurgy has been recently impacted by the emergence of high-entropy alloys (HEAs). 
In contrast to conventional alloys centering around one primary element with minor amounts of other elements,
HEAs mix five or more elements at equal or near-equal compositions often in a single crystalline phase\cite{Cantor2004, Yeh2004}.
The seemingly surprising stabilization of multicomponent alloys against 
the formation of multiple phases and intermetallics (IMs) 
has been associated with the high configurational entropy\cite{Yeh2004} 
among other important factors\cite{Miracle2017}.
HEAs can exhibit unusual properties\cite{Miracle2017,George2019} from 
exceptional toughness at cryogenic temperatures\cite{Otto2013},
to outstanding combination of strength and ductility\cite{Gludovatz2014,Li2016},
high damage tolerance\cite{Zhang2015} and corrosion resistance\cite{Qiu2017}. 
While HEAs have first been mainly studied as structural materials, 
the field is now expanding to other areas such as electrocatalysis\cite{Sun2021}, 
thermoelectrics\cite{Zhang2018} and energy storage.\cite{Berardan2016,Sarkar2018,Wang2019,Zhao2020,Lun2021,Ma2021} 
This is happening while the concept of high entropy stabilization is 
extended beyond metallic alloys with the development of high-entropy oxides and ceramics\cite{Rost2015,Oses2020}.
 
HEAs enjoy a vast compositional space. For equimolar quinary alloys, 
there are \num{658008} candidates resulting from the combination of 40 elements.
Yet, only a limited number of equimolar quinary single-phase HEAs have been observed experimentally over the last decade. 
Computational approaches are called upon to understand the driving force towards the formation of HEAs
and ultimately to accelerate the discovery of new HEAs with specific properties.
Indeed, very limited regions in the compositional space have been explored 
and experimental screening alone would be formidable.

Numerous computational methods have been developed to predict the stability of single-phase solid solutions. 
Early models follow the Hume-Rothery theory\cite{HumeRothery1935,HumeRothery1969} 
and rely on simple descriptors such as atomic radius mismatch and tabulated mixing enthalpy to induce the empirical rules
for the formation of multicomponent solid solutions\cite{Yang2012,Guo2013,Wang2015,Gao2017}.
More sophisticated models additionally take into account the free energy of IM compounds\cite{Troparevsky2015,Senkov2016,King2016},
but are still oversimplified in that the IM phases is hypothetical and 
different definitions of the competing IM phase can lead to diverging predictions\cite{Gao2017,Li2020}.
The CALPHAD (calculation of phase diagrams) method has been used to determine the phase formation of HEAs\cite{Senkov2015,Senkov2015a,Chen2018}
although reliable thermodynamic databases are currently limited to a small number of elements\cite{Gorsse2018}.
The application of machine learning (ML) techniques to HEAs is also 
on the rise\cite{Islam2018,AbuOdeh2018,Wen2019,Kostiuchenko2019,Huang2019,Pei2020,Ha2021,Lee2021}.
ML methods typically make use of the empirical descriptors already known to the existing single-phase solid solutions,
and the relatively small training samples make extrapolating to less studied chemistry a bit hazardous.

First-principles methods offer unbiased insights to the thermodynamic properties of HEAs. 
These methods do not suffer from the fundamental issue with ML models when used for extrapolating to chemical regions that are not well experimentally explored.
Enthalpies obtained from density functional theory (DFT) calculations have already been used
in some semi-empirical models\cite{Troparevsky2015} and CALPHAD\cite{Gorsse2018}.
However, a full \textit{ab initio} treatment of HEAs either involves supercells that are sufficiently large 
to accommodate the configurational disorder\cite{Gao2016}, 
or relies on statistical methods such as cluster expansions\cite{Sobieraj2020,Zhang2020,Lun2021}. 
Either method poses a challenge due to the computational complexity and, when directly applied, is not suitable 
for high-throughput computational screening of HEAs.

In this work, we search possible single-phase HEAs among all equimolar quinary compositions from the combination of 40 metallic elements
that are commonly used in alloys (Supplementary Fig.~1).
This high-throughput computational screening is made possible by the use of a regular solution model\cite{Takeuchi2005,Takeuchi2010}
for which the interactions are described by binary terms and are obtained with DFT calculations.
The thermodynamic stability of HEAs is determined by the Gibbs free energies of the system in solid solutions
against those of the competing phases including IMs.
Our computational model identifies 
\num{30201} equimolar quinary HEAs, 
with the majority (75\%) being BCC. 
Our work offers thus a map of the single-phase high entropy alloys indicating which chemistries favor the formation of these alloys. 
We identify that a high melting point ($T_{m}$) of the elements is among the most important driving factor the formation of HEAs. 
In addition, some outlier elements, such as Al and Zn, are found to form HEAs easily despite their low melting point.
We use our model to predict two equimolar single-phase HEAs, 
namely the BCC \ce{AlCoMnNiV} and the FCC \ce{CoFeMnNiZn} and we confirm experimentally their existence.  
The discovery of a BCC alloy and a FCC alloy analogous to the Cantor alloy (\ce{CoCrFeMnNi}) but with an unusual element Zn
is a compelling demonstration of how our thermodynamic model can suggest chemistries and new avenues to the development of HEAs.

\section*{Results}
\subsection*{Computational model and validations}
\begin{table*}
\caption{\label{tab:model}Predictive metrics of the present thermodynamic model in comparison with various empirical rules (ERs) 
         and free-energy models (FEMs). 
         For the present and the two FEMs, temperature $T$ is chosen such that the best accuracy can be attained
         with the specific model.}
\begin{ruledtabular}
\begin{tabular}{lcccccccc}
  & Present ($T$=1350 K) & ER1\cite{Yang2012} & ER2\cite{Guo2013} & ER3\cite{Wang2015}
  & ER4\cite{Singh2014} & FEM1\cite{Troparevsky2015} ($T$=1500 K) & FEM2\cite{Senkov2016} ($T$=1350 K)\\
\hline
TPR & 70 &  95 & 66 & 66 & 63 & 58 & 58 \\
FPR & 21 &  80 & 54 & 54 & 49 & 33 & 48 \\
Accuracy &  74 & 60 & 57 & 57 & 57 & 62 & 55 \\
\end{tabular}
\end{ruledtabular}
\end{table*}
The thermodynamic stability of an alloy at a given temperature and pressure results from the competition 
between the Gibbs free energy of all competing phases. 
Here we use a regular solution model for all solid solution phases. 
The regular solution model combines an enthalpy model with a quadratic dependence in composition with 
an ideal configurational entropy (see Methods). 
We have previously shown that binary enthalpic interactions are sufficient to reproduce 
the mixing enthalpy of higher component (quaternary and quinary) random solid solutions\cite{Bokas2021}. 
Within this model, the Gibbs free energy of any random solid solution can be computed from 
a series of binary interactions that can be fitted, for instance on DFT.
We have built such a database for a set of 40 elements using the special quasirandom structure approach (SQSs)\cite{Zunger1990}. 
Using these regular solution Gibbs free energies, the competition between all phases can be assessed with the convex hull construction 
which directly compares the free energy of a phase versus any linear combination of its subsystems. 
We additionally include competition from ordered, IM phases up to ternaries 
as provided by the \textsc{Aflow} database\cite{Curtarolo2012}. 
We assume no configurational entropy for the IMs as 
they have well-defined occupancy of the lattice, thus bearing no configurational degree of freedom.
More computational details are provided in the Methods section and our database of 
regular solution enthalpic parameters are available via an online repository\cite{zenodo}.
The convex hull construction can be used to compute if an equimolar solid solution is stable for a given combination of elements. 
Our enthalpic model refrains from using any experimental parameters and is therefore fully \textit{ab initio}.
In addition to thermodynamic stability, our model informs the specific phase for stable HEAs
or the decomposed phases for unstable ones.

The key parameter governing the phase stability assessment in the present study is the temperature $T$ at which the free energy is determined. In experiment, this temperature can be the synthesis temperature or an annealing temperature. Our model can be used to predict if a given equimolar composition will form a single-phase solid solution or will decompose in several other phases. Naturally, higher temperatures favor the entropic contribution and stabilize the single-phase solid solution. We note that if an alloy can be made as a random single-phase solid solution at a high temperature, it will be likely to be retained when quenched. 
Prolonged annealing at intermediate temperatures may lead to phase decomposition, rendering the single-phase HEAs unstable
as is the case with the Cantor alloy\cite{Cantor2004,Schuh2015,Otto2016}. 
Nevertheless, the solid-solution phase formed at high temperatures can still be retained at room temperature following normal cooling rates\cite{Otto2016}. So, the requirement to form a high entropy single-phase equimolar solid solution will be here to show a (reasonable) temperature at which this single-phase is predicted to be stable according to our thermodynamic model.

To validate the predictive power of our model, we use 134 equimolar quaternary and quinary alloys that have been synthesized and 
structurally characterized experimentally\cite{Gao2016a,Gorsse2018a,Borg2020}. This data set includes 73 single-phase HEAs (Supplementary Table~2) and 61 multi-phase alloys (Supplementary Table~3). We used our Gibbs free energy model to see if a single-phase or multi-phase is predicted and if its agrees or disagrees with the experimental report. As the experimental data uses different heat treatments, syntheses and annealing temperatures, we have used a series of temperatures ($T$) in our model from 800 to 1600 K as typically used in the processing of metallic alloys. 
The predictive power of the model is assessed by the true positive rate (TPR) and the false positive rate (FPR),
which are defined by the rate of predicted single-phase solid solutions from the 73 single-phase HEAs 
and from the 61 multi-phase alloys, respectively.
A higher TPR indicates that the model is better at predicting true single-phase solid solutions,
while the model with a higher FPR is considered to be overinclusive for single-phase solid solutions
and is thus less reliable for predicting multi-phase alloys.
The overall accuracy is determined by a combination of TPR and FPR as 
$\left[ \text{TPR}\times 73+(100-\text{FPR}) \times 61 \right]/134$.
We find our model attaining a predictive accuracy of 74\% at the optimal $T=1350$~K 
(Table~\ref{tab:model} and Supplementary Table~2).
Specifically, our model predicts correctly 70\% of the single-phase HEAs
and 79\% of the multi-phase alloys, suggesting that the model performs equally 
well regardless of the phase.

{To put our model in perspective, we apply four empirical rules (ERs)\cite{Yang2012,Guo2013,Singh2014,Wang2015}
and two free-energy models\cite{Troparevsky2015,Senkov2016} developed previously to the same alloy dataset.
In addition to the common criteria such as mixing enthalpy and entropy, 
the ERs rely mainly on atomic size mismatch  
whereas the two FEMs account for the formation of competing IMs (Supplementary Table~1).
Notably, our model consistently outperforms all the ERs and the FEMs as shown in Table~\ref{tab:model}.
Only the ER1\cite{Yang2012} achieves a higher TPR,
but this comes at the cost of a markedly high FPR, showing that the model is strongly skewed towards the formation of single-phase solid solutions.
The two FEMs are less predictive than our model irrespective of the temperature (Supplementary Table~2).
The supremacy of our model is further made apparent by plotting the TPR vs FPR 
analogous to the receiver operating characteristic (ROC) analysis (Supplementary Fig.~2) as our model overall provides always better TPR and FPR than other models irrespective of the temperature used.
It is noteworthy that Al-containing HEAs are normally rejected by the ERs (except ER1) and FEMs 
mainly because the mixing enthalpies involving Al can be very negative and
most models assume that solid solutions are unlikely to form if the mixing enthalpy is too strong.
Nonetheless, such Al-containing HEAs are correctly predicted with our model by a large extent.
A full breakdown of the results are given in Supplementary Table~3.}

In addition to phase stability, 
a reasonable accuracy (74\%) is achieved for predicting the structure of HEAs (Supplementary Table~4).
This is comparable to the valence-electron concentration (VEC) model\cite{Guo2011}
although the original VEC model does not account for the HCP structure.
If we make no distinction between FCC and HCP and treat the two simply as close-packed (CP), 
the accuracy is further improved to 84\%.
Therefore, our model is more capable of predicting if a HEA is BCC or CP (FCC or HCP) as we will discuss more later.
In summary, our regular solution model is at least as effective as previous models and 
rely on a physically-driven approach based on DFT.
It will therefore likely extrapolate better than approaches trained on a small dataset. 

\subsection*{Chemical map of high-entropy alloys}
\begin{figure*}
\includegraphics[width=16cm]{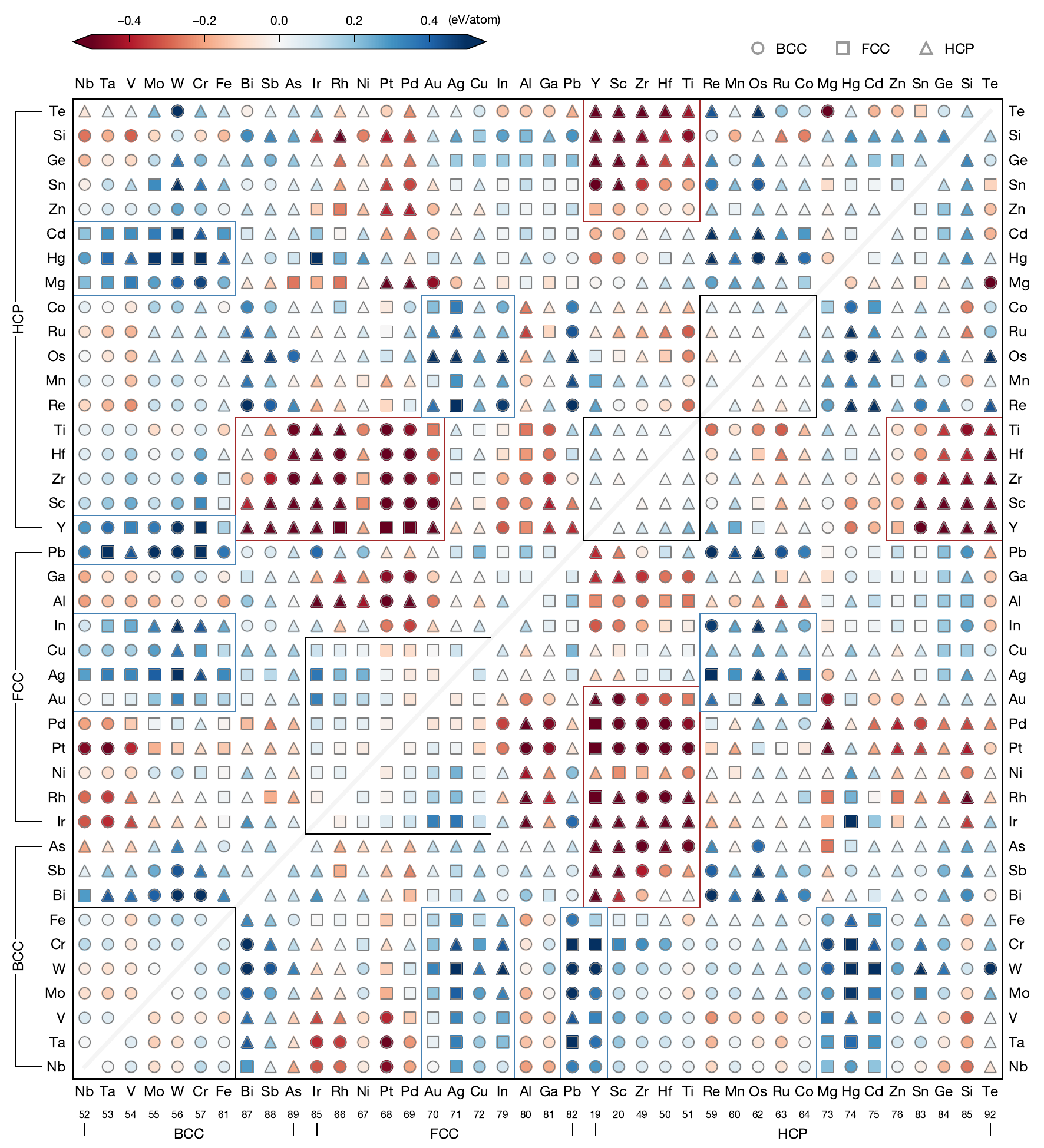}
\caption{\label{fig:matrix}{Map of formation enthalpy for binary solid solutions,
as represented by SQS, obtained from DFT calculations.
The formation enthalpy ($\Delta H^\text{f}$) is determined with respect to the ground-state elemental phases.
The 40 elements are grouped by their lowest energy structure at 0~K (BCC, FCC, or HCP) 
and are sorted according to Pettifor’s Mendeleev numbers\cite{Pettifor1984}.
Groups of elements mixing in the same crystal structure are shown by the black blocks.
Groups of elements that strongly favor mixing ($\Delta H^\text{f} < -0.2$ eV/atom) 
are highlighted by the red blocks, 
whereas those strongly disfavoring mixing ($\Delta H^\text{f} > 0.2$ eV/atom)
is highlighted by the blue blocks.}
}
\end{figure*}

{We now set out to navigate the huge chemical space of quinary alloys from the combination of 40 elements.
To set ourselves in the best scenario for the formation of single-phase solid solutions, 
we choose here $T=0.9T_m$ so that the entropic mixing contribution is maximized.
As outlined before, we consider that HEA formed at high temperature can be quenched to room temperature preserving their single-phase nature.
When tested with the same validation dataset, 
our model at 0.9$T_m$ shows a high TPR of 84\% but a FPR of 62\%. The high TPR is more relevant in the current context of finding new HEAs.}

Applying our model to the \num{658008} possible equimolar quinary alloys,
{we find \num{30201} potentially stable HEAs at
 $T=0.9T_{m}$, which amounts to 4.6\% of the quinary candidates. 
The majority of the stable HEAs (74\%) are found in the BCC structures (Supplementary Table~5).}
Among the \num{7570} CP alloys, the model suggests a large amount of HCP alloys which disagrees with experimental knowledge\cite{Zhang2014}. 
As noted above, our model is less capable of discriminating between HCP and FCC in view of their small difference in energy 
(17~meV/atom when averaged over 75 known HEAs). 
Moreover, we tend to overestimate the stability of the HCP structures as the model does not take into account vibrational entropy 
which in general favors FCC vs HCP at high temperature. 
We estimate the effect of vibrational entropy using the CALPHAD entropy data for a set of 26 elements\cite{Chen2018}. 
On average the vibrational entropy ($-S^\text{vib}$) of the HCP (BCC) structure is 24 (18)$\times$10$^{-3}$~meV$\cdot$K$^{-1}$$\cdot$atom$^{-1}$ 
higher than that of the FCC. 
This stabilization of the FCC vs HCP with temperature has been observed, for instance, 
in the Cantor alloy both experimentally\cite{Zhang2017,Tracy2017} and computationally\cite{Ma2015}. 
{We note that the observed trend also applies to the phase stability analysis at lower temperatures
albeit the predicted number of stable HEAs being reduced (Supplementary Table~5).}
The data on the thermodynamic stability of the \num{658008} quinary alloys are accessible via an online repository\cite{zenodo}.

One of the important factors driving the formation of quinary HEAs is the possibility for the five elements to enthalpically favorably mix in the solid solution.
In our binary regular solid solution model, this is evaluated by the mixing enthalpy $\Delta H^\text{mix} = \frac{4}{25}\sum_{i,j>i} \Delta H^\text{mix}_{i,j}$
for a quinary equimolar alloy where $\Delta H^\text{mix}_{i,j}$ refers to the binary mixing enthalpy for the solid solution. 
The mixing enthalpy of a quinary solid solution is then the results of a sum over all mixing enthalpies of pair combinations of elements. 
For instance, the mixing enthalpy of the Cantor alloy (CoCrFeMnNi) is the result of the different combinations of binaries
(Co--Cr, Co--Fe, Co--Mn, etc.).
Figure~\ref{fig:matrix} gives an overview of the tendency to mix for all pair combinations of the 40 elements.
For each pair of elements, we plot the enthalpy of formation (i.e.\ with respect to the elemental phase in its lowest energy structure)
and the crystal structure in the ground state at 0 K for the binary solid solutions. 
This plot takes into account not only the mixing on a specific lattice (BCC, FCC, HCP) but also the competition between these lattices. 
Figure~\ref{fig:matrix} gives a direct look at what pairs of elements will favor or disfavor mixing. 

Groups of pairs of elements strongly favoring the formation of solid solutions have been indicated by 
red blocks in Fig.~\ref{fig:matrix}. 
For instance, FCC noble metals (Ir, Rh, Ni, Pt, Pd, Au) strongly mix with some HCP transition metals (Ti, Hf, Zr, Sc, Y),
and the same set of HCP transition metals mix favorably with main-group elements (Te, Si, Ge, Sn, Zn).
By contrast, we have also indicated with blue blocks highlighting the regions of disfavorable mixing. 
BCC refractory elements do not mix with a large group of CP elements (Cd, Hg, Mg, Y, Pb, In, Cu, Ag, Au).
Among the other element pairs strongly disfavoring the mixing are some FCC elements (Au, Ag, Cu, In) with HCP elements (Co, Ru, Os, Mn, Re).

\begin{figure}[t]
\includegraphics{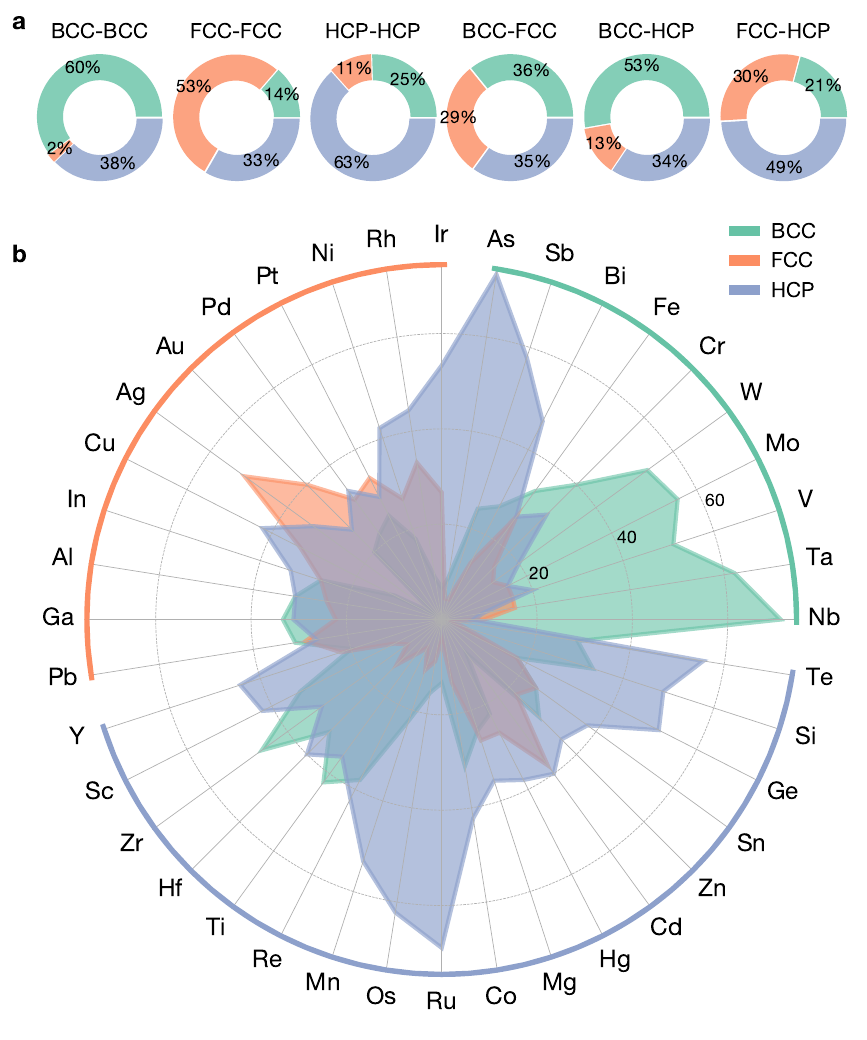}
\caption{\label{fig:elem_phase}Predicted structural preference of binary solid solutions. \textbf{a} Structural preference of binary solid solutions formed through the mixing of two elements of specific ground-state structures (S$_1$--S$_2$) where S$_{1,2}$ refer to either BCC, FCC, or HCP. The statistics are based on the 40 candidate elements.
         \textbf{b} Ground-state structures of binary solid solutions summarized per constituent element. The values refer to the proportion 
         of the structure among all structures found in the binary solid solutions containing the specific element. 
}
\end{figure}

Our map can also be used to understand the prevalence of the stable HEAs 
in the BCC structure. More than 74\% of the HEA form in the BCC structure while only 25\% of the elements are BCC.
While the crystal structure is likely to be maintained as a result of mixing two isostructural elements, 
it is not uncommon for certain elements to end up in solid solutions with a different structure than their elemental ones. 
Figure~\ref{fig:elem_phase}{a} indicates the statistics for the mixing of different structures in binary solid solutions.
Remarkably, the BCC--FCC/HCP mixing leads to the majority of solid solutions being BCC, 
thereby explaining the large number of BCC HEAs.
The elementwise analysis in Fig.~\ref{fig:elem_phase}{b} shows the preference for specific structures in binary solid solutions depending on the elements. 
The refractory elements (Nb, Ta, V, Mo, W) are remarkable for their strong tendency towards the formation of BCC binary solid solutions. 
These BCC refractory elements, when intermixed, consistently retain the BCC structure.
FCC elements such as Al, Ga, or Pb favor as much BCC as CP structures despite their CP nature as elements.
A similar effect is seen with HCP elements such as Zr, Hf, Ti, and Re.
On the other hand, Mn, Os, and Ru are among the elements that are outstanding HCP formers,
whereas Ag and Au are more likely to be found in FCC solid solutions. 

\begin{figure}[t]
\includegraphics{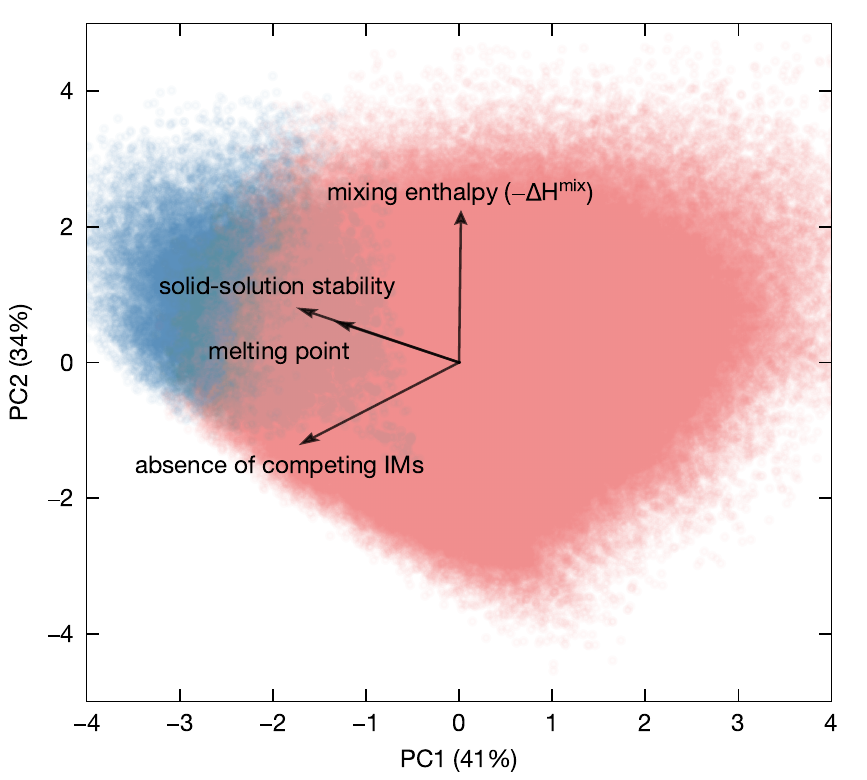}
\caption{\label{fig:pca}Principal component analysis (PCA) biplot depicting the correlations between phase stability,
         melting point, mixing enthalpy, and competing IMs. 
         The degree of competing IMs is assessed by the change in the energy above hull of the hypothetical HEA upon the introduction of IMs. 
         The scores of the variables for the first two principal components (PC1 and PC2) are shown by the scattered points (blue for stable alloys and light red for unstable ones).
         The two components account for 75\% of the explained variance ratio. The loadings are scaled by a factor of 2.5 for clarity.
}
\end{figure}

Apart from mixing energetics, the formation of HEAs is also determined by melting point and IM formation. A high melting point will offer the possibility to entropically stabilize the solid solution through a high synthesis temperature. This is especially important when strong IM formation could destabilize the solid solution.
To clarify the statistical importance of these three factors, 
we perform a principal component analysis (PCA) on the quinary HEAs dataset (Fig.~\ref{fig:pca}).
By projecting the multiple variables onto a lower dimensional space, the PCA is instructive in identifying the correlations
among the variables.
The two variables are more (anti)correlated if the loading vectors are more (anti)parallel,
whereas they are uncorrelated if the loadings become orthogonal.
The degree of stability is defined by the energy above convex hull if the HEA is unstable or by the inverse
energy above hull (i.e.\ the equilibrium reaction energy) otherwise.
The effect of IMs is quantified by the change in the energy above hull 
upon the introduction of binary IMs.
The stability of the alloys can be clearly discriminated from the PCA scores.
It is apparent that the stability is mainly correlated with the melting point
and, to a lesser extent, with the competing IMs and the mixing enthalpy of solid solutions.
The formation of HEAs is therefore strongly driven by the possibility of a high 
synthesis temperature whereby the entropic stabilization effect is amplified.

\begin{figure}[t]
\includegraphics{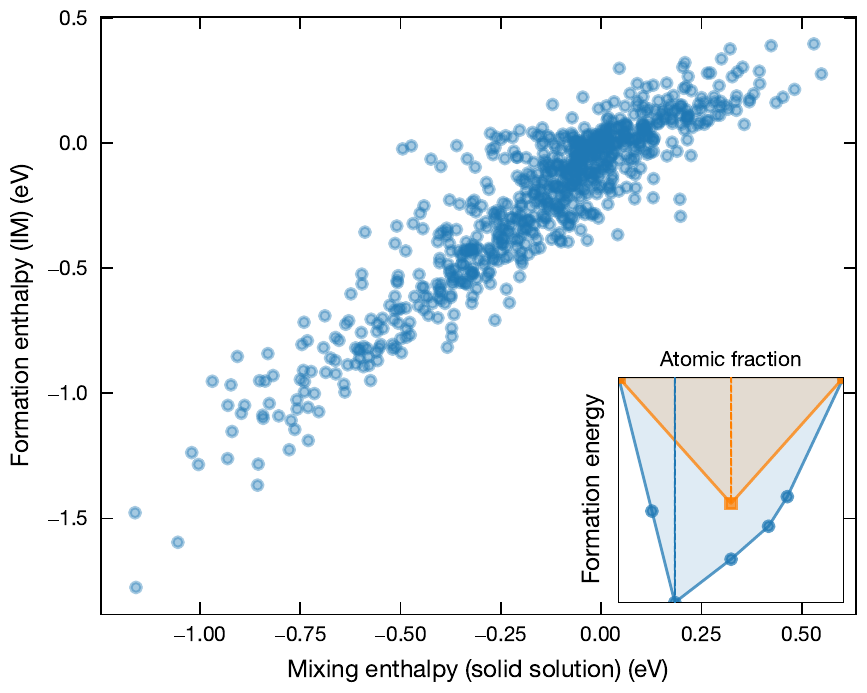}
\caption{\label{fig:im_ss}{Lowest formation enthalpy of IMs 
        vs mixing enthalpy of equimolar solid solutions for binary systems.
        The two quantities are illustrated by the schematic of
        the convex hull for solid solutions (orange) and IMs (blue) in the inset.
        }
}
\end{figure}

The milder effect of the mixing enthalpy can be surprising at first but is rationalized by its correlation with IM formation. 
Elements that tend to mix strongly as a solid solution are also more likely to form ordered IM phases that in turn destabilize the solid solution. 
This was hinted previously by Senkov and Miracle\cite{Senkov2016}. 
Taking all binary systems from the combination of the 40 elements, 
we explicitly show the linear correspondence between the formation enthalpy of IMs 
and the mixing enthalpy of solid solutions in Fig.~\ref{fig:im_ss}.
Following Fig.~\ref{fig:matrix}, we build a matrix indicating the competition between IMs 
and solid solutions for any given pair of the 40 elements (Supplementary Fig.~5). 
Interestingly, the Cantor alloy contains elements that tend to mix mildly together, not too favorably, not too disfavorably. 
While this could seem at first sight to be detrimental for HEA formation, 
this mild mixing in solid solution correlates with weak competition from IMs
(Fig.~\ref{fig:elem_phase} and Supplementary Fig.~6).

\begin{figure}[t]
\includegraphics{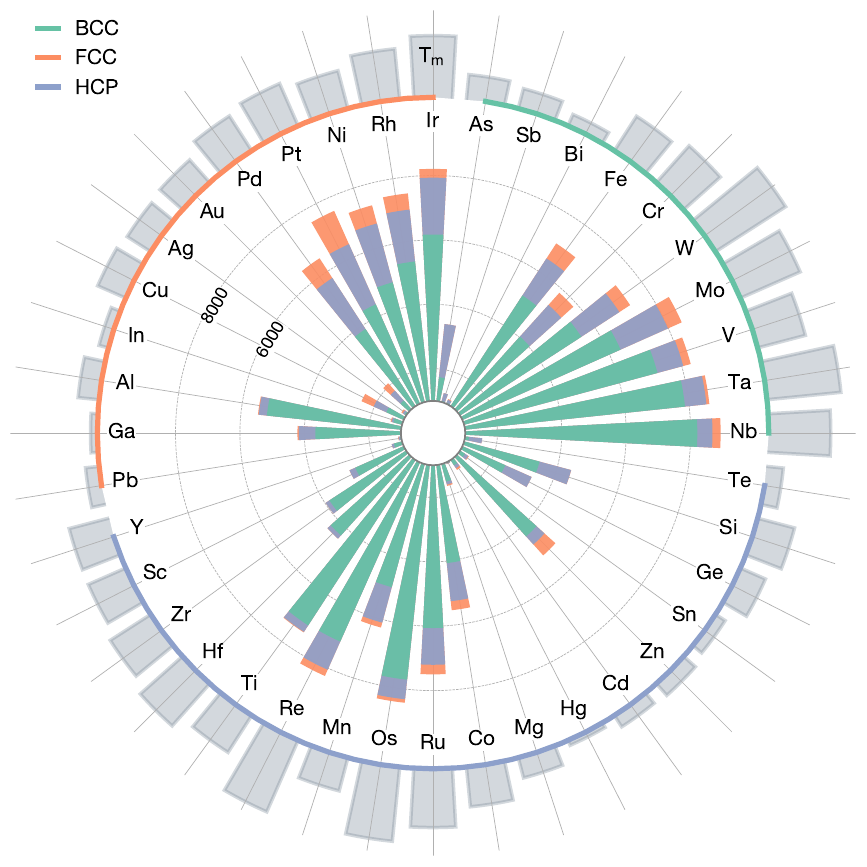}
\caption{\label{fig:elem_5}Number of stable quinary solid solutions per constituent element. The elements are grouped according to their ground-state structures.
        The elemental melting points are indicated by the height of the gray bars on the outskirt.
}
\end{figure}

To probe how chemistry influence the stability of HEAs from the combination of melting temperature and mixing enthalpy, 
we summarize in Fig.~\ref{fig:elem_5} the elemental distribution of our predicted HEAs. 
We also add the elemental melting points.
The prevalence of BCC HEAs is again remarkable. 
In addition to the mixing effect discussed previously where BCC is favored when mixing elements even if they alone form in CP structures, 
BCC is also favored in HEAs because high melting point is often found among the BCC refractory metals.
{In fact, nearly 80\% of the predicted quinary HEAs contain at least one refractory element 
(Cr, W, Mo, V, Ta, or Nb) (Supplementary Table~5), and 77\% of these HEAs are stabilized in the BCC structure.}
Experimentally, a large number of refractory HEAs have been identified 
since the work of Senkov~\textit{et al.}\ in 2010\cite{Senkov2010}
and they form the main body of single-phase HEAs as it is clear from the collection of equimolar HEAs 
(Supplementary Table~3).
The CP HEAs are largely formed by the noble FCC (Pd, Pt, Rh, Ir) and HCP elements (Re, Os, Ru), 
all of which have a relatively high melting point (1900--3400~K).
Other noticeable CP HEA formers include Ni, Ti, Mn, and Co.
While the formation of HEAs is overall favored by elements with a high melting point,
we note two elements, namely Al and Zn, that are outliers to this melting-point rule.
Despite the low melting point of Al (933 K) and Zn (692 K), the two elements can be found in a good amount of HEAs. 
We rationalize this as Al and Zn mix easily with elements of a higher melting point near 2000 K (Supplementary Fig.~7a).
This is in line with the average melting point of the predicted HEAs containing Al (2100~K) and Zn (1900~K).
An inverse trend is observed with the Cu, Ag, and Au, 
which form considerably fewer HEAs than the other elements with a comparable melting point ($\sim$1300~K). 
These group-11 elements are known to behave differently than many transition metals and their miscibility 
is low with elements of higher melting point (Supplementary Fig.~7b).

{The analysis hitherto underscores the difficulty in finding HEAs that will be thermodynamically stable at low temperatures
as the single-phase HEAs formed by entropic effect at high temperatures would instead be metastable, often destabilized by competition from IMs.
Among the few alloys that our model predicts to be stable at 0.6$T_m$ are a series of Sn and Cd alloys of low melting point
(such as \ce{CdGaInMgSn}, \ce{AgCdGaMgSn}, and \ce{GaInMgPdSn}),
for which the formation is either driven by a weak mixing enthalpy of solid solutions in the absence 
of strong competition from IMs (\ce{CdGaInMgSn} and \ce{AgCdGaMgSn})
or stabilized by the favorable mixing enthalpy (e.g., Pd--$X$) despite 
the strong competition from IMs (\ce{GaInMgPdSn}) (Supplementary Figs.~4 and 6).}

\subsection*{Discovery and synthesis of \ce{AlCoMnNiV} and \ce{CoFeMnNiZn} HEAs}

HEAs containing Al are among the first synthesized HEAs\cite{Yeh2004} and 
exhibit intriguing phase-dependent strength-ductility properties\cite{Wang2012,He2014,Luo2019,Yao2020}.
According to our model, the \ce{AlCoMnNiV} BCC solid solution shows an inverse energy above hull of $-0.3$~meV/atom 
at 90\% of the estimated melting point ($T_{m}=1626$ K)
despite its strong mixing enthalpy of $-290$~meV/atom, indicative of a subtle competition from the IMs.
Some Al--$X$ pairs ($X$=Co, Ni) indeed strongly favor the IM phase with respect to the solid solutions (Supplementary Fig.~5).
The BCC solid solution of \ce{AlCoMnNiV} is about 30~meV/atom more stable than the CP
due to the presence of the three strong BCC formers (Al, Mn, and V).
Given its moderate melting point and the strong competing IMs, 
the new BCC HEA \ce{AlCoMnNiV} is an interesting test case for validating our model.
As with many other Al-containing HEAs, most empirical models (except ER1) predict \ce{AlCoMnNiV} unstable in single phase. 
This composition has also to our knowledge never been reported in the experimental literature.

\begin{figure*} 
\includegraphics{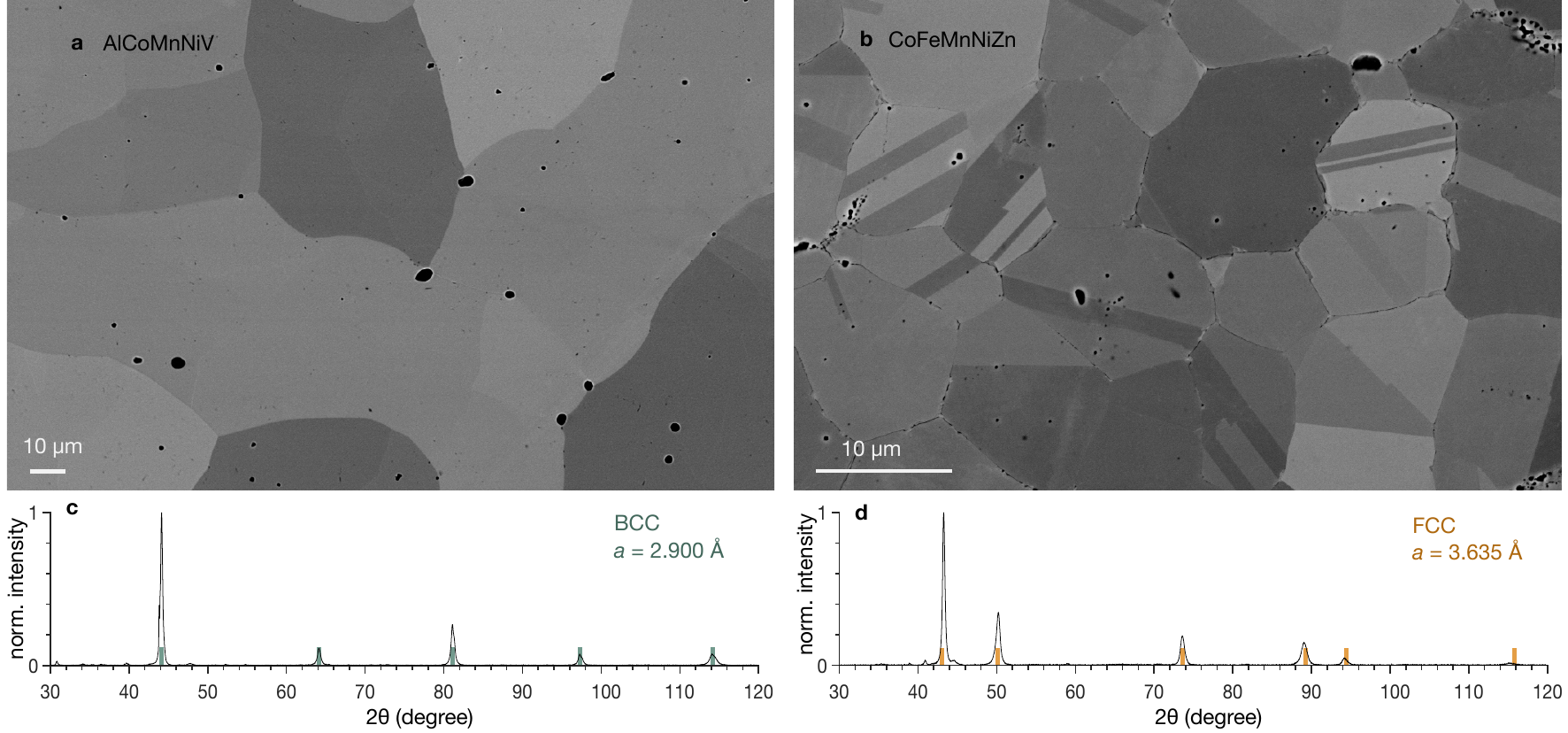} 
\caption{\label{fig:sem_xrd}{Characterizations of the synthesized BCC \ce{AlCoMnNiV} and FCC \ce{CoFeMnNiZn}.
\textbf{a} and \textbf{b} Microstructures of the two HEAs characterized by (BSE) SEM.
\textbf{c} and \textbf{d} XRD spectra of the two HEAs confirming the predicted structures. The vertical markers indicate the reflections for  BCC and FCC structures.}
}
\end{figure*}

To experimentally validate our prediction, \ce{AlCoMnNiV} was synthesized by arc-melting. 
Figure~\ref{fig:sem_xrd}a shows the backscattered electrons (BSE) scanning electron microscopy (SEM) image of the microstructure of the alloy after heat treatment where single phase, equiaxed grains are visible. 
The black spots are porosities due to the solidification. 
X-ray diffraction (XRD) analysis confirmed that this HEA presents a BCC structure. 
Figure~\ref{fig:sem_xrd}c shows the XRD pattern with the reflection associated to the BCC structure with a lattice parameter of $2.9$~\AA, 
highlighted in blue, matching each peaks. The chemical homogenity is confirmed by energy dispersive spectrometry (EDS) (Supplementary Fig.~8).

Compared to BCC HEAs, it is much more difficult to find HEAs stabilized in the FCC and HCP structures.
Of all possible combination of quinary equimolar alloys, we predict that only 1\% form CP alloys.
CP HEAs exhibit some unique characteristics.
For example, FCC HEAs are considerably more ductile than BCC ones especially at low temperatures\cite{Joseph2017}.
Bearing in mind elemental cost, 
our model points to a series of cost-effective Zn-containing HEAs with a potential FCC structure.
Here we choose the FCC \ce{CoFeMnNiZn} as it is closely related to the Cantor alloy.
While alloys with the same set of principal elements have been proposed\cite{Suksamran2017,Tapia2018},
the phase of the equimolar \ce{CoFeMnNiZn} has yet to be characterized.
\ce{CoFeMnNiZn} shows a weak mixing enthalpy of $-80$~meV/atom and an inverse energy above hull of $-30$~meV/atom
at 90\% of the melting point ($T_m=1504$~K).
These values are in close agreement with those for the Cantor alloy, suggesting that the 
balance between the IMs and solid solutions is not disrupted by substituting Cr with Zn
(Supplementary Fig.~6).
Despite Zn being a low melting-point element, the absence of strong forming IMs leads to a high stability
of the \ce{CoFeMnNiZn} solid solution down to 900~K.
In fact, our model indicates that Zn can be used to substitute any element of the Cantor alloy while still 
forming the FCC single-phase HEAs,
although \ce{CoFeMnNiZn} is predicted to be the most stable one. 
We note that Zn is not commonly used in HEAs but has been recently discussed in the field of biodegradable alloys\cite{Yang2020}.


We confirm experimentally the prediction of the model and synthesize a \ce{CoFeMnNiZn} single-phase solid solution. 
Annealing twins, indicative of a FCC structure, are clearly visible in several grains in the microstructure shown in Figure~\ref{fig:sem_xrd}b.  
This is confirmed by the XRD pattern of \ce{CoFeMnNiZn} in Fig.~\ref{fig:sem_xrd}d 
which is indexed as FCC with a lattice parameter of $3.635$~\AA. 
The black spots observed in Fig.~\ref{fig:sem_xrd}b corresponds to either MnO oxide particles or porosities (Supplementary Fig.~9). 
Mn is prone to oxidation whereas Zn has a low boiling point, limiting the processing routes available. 
It is worth mentioning that processing this alloy was a more arduous task than for the previous one due to 
the sensitivity of Mn to oxidation and the low boiling point of Zn.

\section*{Discussion}
By accounting for the configurational entropy in addition to the mixing enthalpy,
our \textit{ab initio} driven thermodynamic model achieves an accuracy up to 74\% for predicting the phase stability of multicomponent alloys,
surpassing existing empirical rules and free-energy models. 
The predictive power of the model could be further improved by the inclusion of vibrational effects, 
magnetic ordering, and short-range ordering~\cite{Shen2021,Gao2017,Feng2017,Esters2021} that have been neglected in the present model.
All of these effects would significantly increase the computational cost and prevent the large scale search we have reported on. 
Nevertheless, alloys suggested by our model could be used to perform a more refined modeling 
including these effects within a possible tiered screening approach for HEA discovery.

{To search for potentially stable single-phase HEAs and 
to elucidate the mechanisms underlying their phase stability,
we have navigated the vast chemical space 
of all quinary equimolar alloys from the combinations of 40 elements using our model.}
Among the \num{658008} quinary alloys, we predict that 5\% of them
can be stabilized in a single-phase solid solution at near-melting point.
The amount of predicted equimolar HEAs corresponds to the theoretical upper limit, 
and is significantly more than what has been reported in the literature.
The predicted HEAs show a strong tendency to form BCC phases, in line with the large body 
of BCC HEAs that have already been identified.
Our model suggests that the prevailing BCC phase originates from a combined effect 
of the high melting of the constituent BCC elements, which are often refractory, 
and a favorable mixing of elements on a BCC lattice.
The high melting point is in fact one of the main driving forces for the single-phase HEA formation.
By that token, many closed-packed alloys, such as the FCC Cantor alloy, are more the exception than the rule
as they normally contain zero to very few refractory elements. 

The map of binary interactions presented here is instructive in rationalizing and predicting
the chemistries that are likely to lead to new HEAs.
The series of Al- and Zn-containing HEAs show that non-refractory and cost-effective HEAs 
with a relatively low melting point can be stabilized by the subtle enthalpic competition 
between IMs and solid solutions.
The successful synthesis of the BCC AlCoMnNiV and the FCC CoFeMnNiZn signals the 
promising application of our current approach towards the quest of new HEAs.

{While our work does not inform properties other than the phase stability, 
additional computational screenings driven by specific desired properties can be envisaged in combination with our thermodynamic model.
In addition, the present approach is applicable to a range of technogically relevant 
temperatures and can readily be applied to alloys deviating from the equimolar composition.}

\section*{Methods}
\subsection*{Binary solid-solution model}
The Gibbs free energy of mixing at temperature $T$ can be expressed as
$\Delta G^\text{mix} = \Delta H^\text{mix} - T\Delta S^\text{mix}$.
Within the binary regular solution model, the enthalpy of mixing $\Delta H_\text{mix}$ of an $n$-component
system can be written as a linear combination of the pair interactions among the constituent elements
\begin{equation}
\Delta H^\text{mix} = \sum_i \sum_{j>i} \Omega_{ij} c_i c_j,
\end{equation}
where $\Omega_{ij}$ are the binary interaction between atoms $i$ and $j$ and the sum runs through all combination of pairs, 
and $c_i$ is the molar fraction of the $i$th element. 
The mixing entropy in a regular solution is the ideal mixing entropy
\begin{equation}
\Delta S^\text{mix}_\text{ideal} = -R\sum_i c_i \text{ln} c_i,
\end{equation}
where $R$ is the gas constant. 
The binary interaction is obtained from the enthalpy of mixing of the binary system as
\begin{equation}
\Omega_{ij} = 4 \Delta H_{ij}^\text{mix} = 4 \left[ E^\text{SQS}_{ij} - \frac{1}{2}(E_i + E_j) \right],
\end{equation}
where $E_{ij}^\text{SQS}$ is the total energy of the binary system represented by the special quasirandom structure (SQS),
and $E_{i}$ is the total energy of the elemental system $i$ in the same lattice as the parent binary structure.
The 16-atom SQS structures for the FCC, BCC, and HCP structures used in this work are generated
by the \textsc{atat} suite of software\cite{vandeWalle2002}, whereby pair (triplet) interaction up to the 6th (3rd) nearest neighbor
are taken into account.
This Gibbs free energy of mixing is defined for a given structural lattice and computed on FCC, HCP and BCC lattice.

For the construction of energy convex hull, the Gibbs free energy of formation is used and can be easily obtained from $\Delta G^\text{mix}$
\begin{equation}
\Delta G^\text{f} = \Delta G^\text{mix} + 2c_ic_j\sum_i\sum_{j>i} \left[ (E_i + E_j) - (E_i^0 + E_j^0) \right],
\end{equation}
where the superscript 0 denotes that the total energy of an elemental system is calculated at its ground-state structure
(as opposed to the same structure as the parent lattice used for calculating enthalpy of mixing).

\subsection*{Competing intermetallics}
{For a given quinary alloy, we search all its binary and ternary compounds using the \textsc{Aflow} 
\textsc{icsd} and \textsc{lib} collections of IMs.
We construct a convex hull of formation energy for the available IMs for a specific binary or ternary composition,
from which the stable compounds and the unstable compounds with an energy above hull of less than 10~meV/atom
are chosen for a refining DFT computation using the same parameters as for the solid solutions.
The 10~meV/atom cutoff threshold takes into account the uncertainties arising from the differences
between the computation parameters (e.g., pseudopotentials, kinetic energy cutoffs, and $\mathbf{k}$-point samplings) 
used by the \textsc{Aflow} dataset and our present calculations.
Supplementary Fig.~3 shows that the threshold of 10~meV/atom suffices to include as many 
stable IMs as the numerical uncertainty of the formation energy is indeed typically within this value.
The calculated IM entries (about 9100 binaries and 7800 ternaries) are then added together with the solid solutions
to the energy convex hull in order to assess the competition from the IMs.}

{While we consider the IMs up to the ternaries, the effect of competing phases is largely
captured by the binary systems. If only the binary IMs are considered in our model,
the number of stable quinary single-phase HEAs would be \num{35608}, i.e.\ about 10\% more than in 
the presence of ternary IMs.
Any higher order IMs would have an negligible effect.}

\subsection*{Density-functional theory calculation}
DFT calculations for the SQSs and IMs are performed within the
Perdew-Burke-Ernzerhof (PBE) generalized-gradient approximation using the \textsc{vasp}
code\cite{Kresse1996,Kresse1999}
The planewave energy cutoff is 500~eV, and a grid density of \num{2000}~$\mathbf{k}$ points per atom is used.
The atomic positions are relaxed until the forces are smaller than 0.02~eV/\AA.
For the BCC and FCC SQSs, the lattice parameter is optimized while the cell shape is intact.
For the HCP SQS, an additional constraint is imposed on the the $c/a$ ratio.
Specifically, we fully optimize the HCP structure for each of the constituent elements and take the
average $c/a$ value for the HCP SQS.
All calculations are spin polarized and initialized with a ferromagnetic configuration.

\subsection*{Processing and analysis of \ce{AlCoMnNiV} and \ce{CoFeMnNiZn}}
\ce{AlCoMnNiV} is prepared by arc melting of pure elements ($>99.9$\% purity). 
A pre-alloy of manganese and nickel is made to avoid Mn evaporation issues. 
Furthermore, the Mn chips are deoxidized prior to casting using 50\% HCl solution. 
The alloy is re-melted at least five times in the arc furnace before being heat treated at 1373~K for 24 hours.

The processing of \ce{CoFeMnNiZn} is more challenging due to the presence of Zn. 
Due to its low melting (693~K) and boiling point (1180~K), arc or induction melting is not appropriate. 
Instead, \ce{CoFeMnNiZn} is obtained by mechanical alloying of pure elements (powders of $>99.9$\% purity). 
Powder pre-alloying is carried out before sintering for densification. 
10-mm-thick pellets are compacted in a die of 12~mm in diameter with of force of 50 kN and heat treated in Ar-filled 
quartz capsule for 5 days at 1073~K followed by 5 hours at 1273~K. 
The heat-treated pellets are then ground and the resulting powder is densified by spark plasma sintering (SPS). 
The sintering is carried in a 30-mm die at 48~MPa and 1123~K with a holding time of 2.5~min. 
The heating cycle is carried out under Ar atmosphere.

Sample preparations follow standard practices starting with mechanical polishing with SiC, followed by 6 and 1~$\mu$m diamond paste polishing. The final step consists in a polishing with a solution containing silica oxide particles in suspension (OPS). The microstructure is characterized by scanning electron microscopy (SEM) at 15~keV. 
Local chemical analysis is measured by energy dispersive X-ray spectrometry (EDS). 
The x-ray diffraction (XRD) with Cu radiation (wavelength of $1.541$~\AA) operated at 30~kV and 30~mA is performed on unetched samples to characterize the crystallography.

\section*{Data availability}
The phase stability analysis of the \num{658008} quinary alloys is available as a csv file at
\url{https://doi.org/10.5281/zenodo.7633180}.
The binary interactions of the 40 elements and the formation energies of the binary and ternary intermetallics
are also available as json files and can be downloaded from the repository.

\section*{Code availability}
The code used in the present work for the phase stability analysis is available at \url{https://doi.org/10.5281/zenodo.7633180}.

\begin{acknowledgments}
The research was funded by the Walloon Region under the agreement No.\ 1610154-EntroTough in the context of the 2016 Wallinnov call. 
Computational resources were provided by the Consortium des \'Equipements de Calcul Intensif (C\'ECI), 
funded by the Fonds de la Recherche Scientifique de Belgique (F.R.S.-FNRS) under Grant No.~2.5020.11 and by the Walloon Region.
The present research benefited from computational resources made available on the {Tier-1} supercomputer of the F\'ed\'eration Wallonie-Bruxelles, infrastructure funded by the Walloon Region under grant agreement No.~117545.
\end{acknowledgments}

\bibliography{main}

\end{document}


\title{\textsc{Supplementary Information} \\ A map of single-phase high-entropy alloys}
       
\author{Wei Chen}
\affiliation{Institute of Condensed Matter and Nanoscicence (IMCN),
Universit\'{e} catholique de Louvain,
Chemin Etoiles 8, 
Louvain-la-Neuve 1348, Belgium}
\author{Antoine Hilhorst}
\affiliation{Institute of Mechanics, Materials and Civil Engineering (IMAP),
Place Sainte Barbe 2, Louvain-la-Neuve 1348, Belgium}
\author{Georges Bokas}
\affiliation{Institute of Condensed Matter and Nanoscicence (IMCN),
Universit\'{e} catholique de Louvain,
Chemin Etoiles 8, 
Louvain-la-Neuve 1348, Belgium}
\author{St\'{e}phane Gorsse}
\affiliation{CNRS, University of Bordeaux, Bordeaux INP, ICMCB,
UMR 5026,
Pessac 33600, France
}
\author{Pascal J.\ Jacques}
\affiliation{Institute of Mechanics, Materials and Civil Engineering (IMAP),
Place Sainte Barbe 2, Louvain-la-Neuve 1348, Belgium}
\author{Geoffroy Hautier}
\affiliation{Institute of Condensed Matter and Nanoscicence (IMCN),
Universit\'{e} catholique de Louvain,
Chemin Etoiles 8, 
Louvain-la-Neuve 1348, Belgium}
\affiliation{Thayer School of Engineering, Dartmouth College,
Hanover, New Hampshire 03755, USA}

\date{\today}
\maketitle
\listoftables
\listoffigures
\clearpage

\begin{table}
\caption{\label{tab:def}Criteria for the formation of single-phase HEAs defined by various empirical rules (ERs) and free-energy models (FEMs).}
\begin{ruledtabular}
\begin{tabular}{lll}
ER1 & $T_\text{m}\Delta S_\text{mix}/|\Delta H_\text{mix}| \geq 1.1, \delta < 0.066$\footnotemark[1] & Ref.~\onlinecite{Yang2012} \\
ER2 & $-11.6 < \Delta H_\text{mix} < 3.2$ kJ/mol, $\delta < 0.066$ & Ref.~\onlinecite{Guo2013} \\
ER3 & $-11.6 < \Delta H_\text{mix} < 3.2$ kJ/mol, $\gamma < 1.175$\footnotemark[2] & Ref.~\onlinecite{Wang2015} \\
ER4 & $\Delta S_\text{mix}/\delta^2 > 9.6$ kJ/mol/K & Ref.~\onlinecite{Singh2014} \\
FEM1 & $-T_\text{ann}\Delta S_\text{mix} < \text{min}(\Delta H_{\text{f},ij}^\text{IM}) < 3.57$ meV/atom & Ref.~\onlinecite{Troparevsky2015} \\
FEM2 & $1+0.4\times T_\text{ann}\Delta S_\text{mix}/|\Delta H_\text{mix}| < \Delta H_\text{f}^\text{IM}/\Delta H_\text{mix}$ & Ref.~\onlinecite{Senkov2016} \\
VEC & FCC if VEC $\geq$ 8, BCC if VEC $\leq$ 6.87, mixed phase otherwise & Ref.~\onlinecite{Guo2011}
\footnotetext[1]{$\delta = \sqrt{\sum_i c_i \left( 1 - \frac{r_i}{\bar{r}} \right)^2}$, where $c_i$ is the atomic percentage of the $i$th component, 
$r_i$ the atomic radius and $\bar{r}=\sum_i c_i r_i$.}
\footnotetext[2]{$\gamma = \frac{\omega_s}{\omega_l}$, where $\omega_{s,l} = 1 - \sqrt{ \frac{(r_{s,l} + \bar{r})^2 - \bar{r}^2}{(r_{s,l} + \bar{r})^2}}$.
$r_{s,l}$ refers to the radius of the smallest and the largest atoms, respectively.}
\end{tabular}
\end{ruledtabular}
\end{table}

\begin{table}
\caption{\label{tab:topt}Model validations for the present and the two free-energy models as a function of temperature (in K).
The true positive rate (TPR) refers to the percentage of the single-phase solid solutions predicted 
from a pool of 73 experimentally known single-phase quaternary and quinary HEAs.
The false positive rate (FPR) refers to the percentage of the single-phase solid solutions predicted 
from a pool of 61 multi-phase quaternary and quinary alloys.
The overall accuracy is defined as $\left[\text{TPR} \times 73+(100-\text{FPR}) \times 61\right]/134$.}
\begin{ruledtabular}
\begin{tabular}{lrrrrrrrrr}
 & \multicolumn{3}{c}{Present} & \multicolumn{3}{c}{FEM1} & \multicolumn{3}{c}{FEM2} \\
\cline{2-4}
\cline{5-7}
\cline{8-10}
$T_{a}$ (K) & TPR & FPR & Acc. & TPR & FPR & Acc. & TPR & FPR & Acc.\\
\hline
800 & 22 & 2 & 57 & 22 & 11 & 52 & 42 & 36 & 52 \\
850 & 26 & 3 & 58 & 26 & 11 & 54 & 42 & 39 & 51 \\
900 & 32 & 7 & 60 & 29 & 15 & 54 & 44 & 39 & 51 \\
950 & 37 & 10 & 61 & 32 & 18 & 54 & 47 & 43 & 51 \\
1000 & 44 & 10 & 65 & 37 & 21 & 56 & 47 & 44 & 51 \\
1050 & 45 & 10 & 66 & 37 & 21 & 56 & 47 & 46 & 50 \\
1100 & 47 & 11 & 66 & 41 & 26 & 56 & 47 & 46 & 50 \\
1150 & 47 & 20 & 62 & 41 & 26 & 56 & 47 & 48 & 49 \\
1200 & 55 & 20 & 66 & 42 & 28 & 56 & 49 & 48 & 51 \\
1250 & 60 & 20 & 69 & 44 & 30 & 56 & 49 & 48 & 51 \\
1300 & 63 & 20 & 71 & 47 & 30 & 57 & 52 & 48 & 52 \\
1350 & 70 & 21 & 74 & 47 & 30 & 57 & 58 & 48 & 55 \\
1400 & 74 & 28 & 73 & 49 & 31 & 58 & 59 & 51 & 54 \\
1450 & 75 & 30 & 73 & 49 & 31 & 58 & 62 & 52 & 55 \\
1500 & 79 & 38 & 72 & 58 & 33 & 62 & 62 & 54 & 54 \\
1550 & 82 & 43 & 71 & 58 & 33 & 62 & 63 & 56 & 54 \\
1600 & 88 & 46 & 72 & 58 & 33 & 62 & 64 & 56 & 55 \\
\end{tabular}
\end{ruledtabular}
\end{table}
\clearpage

\begin{longtable*}{llrrcccccccc}
\caption{\label{tab:stable}Predicted phases of the 134 experimentally confirmed equimolar quaternary 
and quinary alloys (compiled from Refs.~\onlinecite{Gao2016a,Gorsse2018a,Borg2020})
by the present approach along with the existing empirical rules (ERs) and free-energy models (FEMs).
The temperature for the present and the two FEMs (FEM1 and FEM2) is set to 1350, 1500, and 1350~K, respectively.
The enthalpy of mixing ($\Delta H^\text{M}_\text{mix}$, in meV/atom) refers to the binary terms taken from 
Ref.~\onlinecite{Takeuchi2010} on the basis of Miedema's scheme \cite{Boer1988}.
The enthapy of formation for the intermetallics ($\Delta H_\text{f}^\text{IM}$, in meV/atom) is defined by 
$\Delta H_\text{f}^\text{IM}=4\sum_{j>i}c_ic_j \Delta H^\text{IM}_{ij}$ 
where the binary term $\Delta H^\text{IM}_{ij}$ is obtained with DFT calculations.
The definitions of the ERs and FEMs are given in Table~\ref{tab:def}.
Single phase and multi phase are denoted by the solid ($\blacksquare$) and the open ($\circledcirc$) markers, respectively.
Correct (incorrect) predictions are highlighted in blue (red).
}\\
\toprule
\multicolumn{1}{l}{\text{Alloy}} &
\multicolumn{1}{l}{$T_m \text{(K)}$} &
\multicolumn{1}{c}{$\Delta H^\text{M}_\text{mix}$} &
\multicolumn{1}{c}{$\Delta H_\text{f}^\text{IM}$} &
\multicolumn{1}{c}{\text{Expt.}} &
\multicolumn{1}{c}{\text{Present}} &
\multicolumn{1}{c}{\text{ER1}} &
\multicolumn{1}{c}{\text{ER2}} &
\multicolumn{1}{c}{\text{ER3}} &
\multicolumn{1}{c}{\text{ER4}} &
\multicolumn{1}{c}{\text{FEM1}} &
\multicolumn{1}{c}{\text{FEM2}} \\
\hline
\endfirsthead
\multicolumn{12}{c}
  {{\tablename\ \thetable{} -- \textit{Continued from previous page}}} \\
\toprule
\multicolumn{1}{l}{\text{System}} &
\multicolumn{1}{l}{$T_m \text{(K)}$} &
\multicolumn{1}{c}{$\Delta H^\text{M}_\text{mix}$} &
\multicolumn{1}{c}{$\Delta H_\text{f}^\text{IM}$} &
\multicolumn{1}{c}{\text{Expt.}} &
\multicolumn{1}{c}{\text{Present}} &
\multicolumn{1}{c}{\text{ER1}} &
\multicolumn{1}{c}{\text{ER2}} &
\multicolumn{1}{c}{\text{ER3}} &
\multicolumn{1}{c}{\text{ER4}} &
\multicolumn{1}{c}{\text{FEM1}} &
\multicolumn{1}{c}{\text{FEM2}} \\
\hline
\endhead
\hline \multicolumn{12}{r}{\textit{Continued on next page}} \\ 
\endfoot
\toprule
\endlastfoot
  AlCoCrNi & 1652 & $-162$ &   $-365$ & $\circledcirc$ & \bingo{$\circledcirc$} & \ooops{$\blacksquare$} & \bingo{$\circledcirc$} & \bingo{$\circledcirc$} & \bingo{$\circledcirc$} & \bingo{$\circledcirc$} & \bingo{$\circledcirc$} \\
  AlCoFeNi & 1560 & $-141$ &   $-435$ & $\blacksquare$ & \ooops{$\circledcirc$} & \bingo{$\blacksquare$} & \ooops{$\circledcirc$} & \ooops{$\circledcirc$} & \ooops{$\circledcirc$} & \ooops{$\circledcirc$} & \ooops{$\circledcirc$} \\
  AlCoNiTi & 1593 & $-346$ &   $-654$ & $\circledcirc$ & \bingo{$\circledcirc$} & \bingo{$\circledcirc$} & \bingo{$\circledcirc$} & \bingo{$\circledcirc$} & \bingo{$\circledcirc$} & \bingo{$\circledcirc$} & \bingo{$\circledcirc$} \\
  AlCrCuFe & 1571 &  $-12$ &   $-109$ & $\circledcirc$ & \bingo{$\circledcirc$} & \ooops{$\blacksquare$} & \ooops{$\blacksquare$} & \ooops{$\blacksquare$} & \bingo{$\circledcirc$} & \bingo{$\circledcirc$} & \bingo{$\circledcirc$} \\
  AlCrFeNi & 1663 & $-138$ &   $-294$ & $\blacksquare$ & \ooops{$\circledcirc$} & \bingo{$\blacksquare$} & \ooops{$\circledcirc$} & \ooops{$\circledcirc$} & \ooops{$\circledcirc$} & \ooops{$\circledcirc$} & \ooops{$\circledcirc$} \\
  AlCrMoNb & 2190 & $-118$ &   $-258$ & $\circledcirc$ & \bingo{$\circledcirc$} & \ooops{$\blacksquare$} & \ooops{$\blacksquare$} & \ooops{$\blacksquare$} & \bingo{$\circledcirc$} & \bingo{$\circledcirc$} & \bingo{$\circledcirc$} \\
  AlCrMoTi & 1988 & $-143$ &   $-289$ & $\blacksquare$ & \bingo{$\blacksquare$} & \bingo{$\blacksquare$} & \ooops{$\circledcirc$} & \ooops{$\circledcirc$} & \ooops{$\circledcirc$} & \ooops{$\circledcirc$} & \ooops{$\circledcirc$} \\
  AlCuNiTi & 1490 & $-257$ &   $-483$ & $\blacksquare$ & \ooops{$\circledcirc$} & \ooops{$\circledcirc$} & \ooops{$\circledcirc$} & \ooops{$\circledcirc$} & \ooops{$\circledcirc$} & \ooops{$\circledcirc$} & \ooops{$\circledcirc$} \\
  AlFeNiTi & 1603 & $-300$ &   $-593$ & $\circledcirc$ & \bingo{$\circledcirc$} & \bingo{$\circledcirc$} & \bingo{$\circledcirc$} & \bingo{$\circledcirc$} & \bingo{$\circledcirc$} & \bingo{$\circledcirc$} & \bingo{$\circledcirc$} \\
  AlHfNbTi & 2033 & $-208$ &   $-322$ & $\circledcirc$ & \ooops{$\blacksquare$} & \bingo{$\circledcirc$} & \bingo{$\circledcirc$} & \bingo{$\circledcirc$} & \bingo{$\circledcirc$} & \bingo{$\circledcirc$} & \bingo{$\circledcirc$} \\
  AlHfTaTi & 2168 & $-214$ &   $-282$ & $\circledcirc$ & \ooops{$\blacksquare$} & \bingo{$\circledcirc$} & \bingo{$\circledcirc$} & \bingo{$\circledcirc$} & \bingo{$\circledcirc$} & \bingo{$\circledcirc$} & \bingo{$\circledcirc$} \\
  AlMoNbTi & 2130 & $-156$ &   $-369$ & $\blacksquare$ & \bingo{$\blacksquare$} & \bingo{$\blacksquare$} & \ooops{$\circledcirc$} & \ooops{$\circledcirc$} & \ooops{$\circledcirc$} & \ooops{$\circledcirc$} & \ooops{$\circledcirc$} \\
  AlNbTaTi & 2229 & $-165$ &   $-287$ & $\blacksquare$ & \bingo{$\blacksquare$} & \bingo{$\blacksquare$} & \ooops{$\circledcirc$} & \ooops{$\circledcirc$} & \ooops{$\circledcirc$} & \ooops{$\circledcirc$} & \ooops{$\circledcirc$} \\
   AlNbTiV & 1952 & $-168$ &   $-291$ & $\blacksquare$ & \bingo{$\blacksquare$} & \bingo{$\blacksquare$} & \ooops{$\circledcirc$} & \ooops{$\circledcirc$} & \ooops{$\circledcirc$} & \ooops{$\circledcirc$} & \ooops{$\circledcirc$} \\
  AlNbTiZr & 1938 & $-222$ &   $-339$ & $\circledcirc$ & \ooops{$\blacksquare$} & \bingo{$\circledcirc$} & \bingo{$\circledcirc$} & \bingo{$\circledcirc$} & \bingo{$\circledcirc$} & \bingo{$\circledcirc$} & \bingo{$\circledcirc$} \\
  CoCrCuFe & 1779 &   $65$ &     $49$ & $\blacksquare$ & \ooops{$\circledcirc$} & \bingo{$\blacksquare$} & \ooops{$\circledcirc$} & \ooops{$\circledcirc$} & \bingo{$\blacksquare$} & \bingo{$\blacksquare$} & \bingo{$\blacksquare$} \\
  CoCrFeNi & 1872 &  $-39$ &    $-45$ & $\blacksquare$ & \bingo{$\blacksquare$} & \bingo{$\blacksquare$} & \bingo{$\blacksquare$} & \bingo{$\blacksquare$} & \bingo{$\blacksquare$} & \bingo{$\blacksquare$} & \bingo{$\blacksquare$} \\
  CoCrMnNi & 1799 &  $-59$ &    $-58$ & $\blacksquare$ & \bingo{$\blacksquare$} & \bingo{$\blacksquare$} & \bingo{$\blacksquare$} & \bingo{$\blacksquare$} & \bingo{$\blacksquare$} & \bingo{$\blacksquare$} & \bingo{$\blacksquare$} \\
  CoCrMoNb & 2398 & $-120$ &    $-80$ & $\circledcirc$ & \bingo{$\circledcirc$} & \ooops{$\blacksquare$} & \bingo{$\circledcirc$} & \bingo{$\circledcirc$} & \ooops{$\blacksquare$} & \ooops{$\blacksquare$} & \ooops{$\blacksquare$} \\
  CoCuFeNi & 1666 &   $53$ &    $-17$ & $\blacksquare$ & \bingo{$\blacksquare$} & \bingo{$\blacksquare$} & \ooops{$\circledcirc$} & \ooops{$\circledcirc$} & \bingo{$\blacksquare$} & \bingo{$\blacksquare$} & \bingo{$\blacksquare$} \\
  CoFeMnNi & 1706 &  $-40$ &   $-100$ & $\blacksquare$ & \bingo{$\blacksquare$} & \bingo{$\blacksquare$} & \bingo{$\blacksquare$} & \bingo{$\blacksquare$} & \bingo{$\blacksquare$} & \bingo{$\blacksquare$} & \bingo{$\blacksquare$} \\
  CoFeNiPd & 1784 &  $-22$ &    $-79$ & $\blacksquare$ & \ooops{$\circledcirc$} & \bingo{$\blacksquare$} & \bingo{$\blacksquare$} & \bingo{$\blacksquare$} & \bingo{$\blacksquare$} & \bingo{$\blacksquare$} & \bingo{$\blacksquare$} \\
   CoFeNiV & 1872 & $-108$ &   $-198$ & $\blacksquare$ & \bingo{$\blacksquare$} & \bingo{$\blacksquare$} & \bingo{$\blacksquare$} & \bingo{$\blacksquare$} & \bingo{$\blacksquare$} & \ooops{$\circledcirc$} & \ooops{$\circledcirc$} \\
  CoFeReRu & 2411 &  $-13$ &    $-47$ & $\blacksquare$ & \ooops{$\circledcirc$} & \bingo{$\blacksquare$} & \bingo{$\blacksquare$} & \bingo{$\blacksquare$} & \bingo{$\blacksquare$} & \bingo{$\blacksquare$} & \bingo{$\blacksquare$} \\
  CoNiRhRu & 2085 &   $-6$ &     $53$ & $\blacksquare$ & \ooops{$\circledcirc$} & \bingo{$\blacksquare$} & \bingo{$\blacksquare$} & \bingo{$\blacksquare$} & \bingo{$\blacksquare$} & \bingo{$\blacksquare$} & \bingo{$\blacksquare$} \\
  CrFeMnNi & 1810 &  $-41$ &    $-64$ & $\blacksquare$ & \bingo{$\blacksquare$} & \bingo{$\blacksquare$} & \bingo{$\blacksquare$} & \bingo{$\blacksquare$} & \bingo{$\blacksquare$} & \bingo{$\blacksquare$} & \bingo{$\blacksquare$} \\
  CrMoTaTi & 2577 &  $-54$ &   $-135$ & $\circledcirc$ & \bingo{$\circledcirc$} & \ooops{$\blacksquare$} & \ooops{$\blacksquare$} & \ooops{$\blacksquare$} & \ooops{$\blacksquare$} & \bingo{$\circledcirc$} & \bingo{$\circledcirc$} \\
  CrNbTaTi & 2540 &  $-47$ &    $-64$ & $\circledcirc$ & \bingo{$\circledcirc$} & \ooops{$\blacksquare$} & \ooops{$\blacksquare$} & \ooops{$\blacksquare$} & \ooops{$\blacksquare$} & \ooops{$\blacksquare$} & \ooops{$\blacksquare$} \\
   CrNbTiW & 2642 &  $-67$ &    $-70$ & $\circledcirc$ & \bingo{$\circledcirc$} & \ooops{$\blacksquare$} & \ooops{$\blacksquare$} & \ooops{$\blacksquare$} & \ooops{$\blacksquare$} & \ooops{$\blacksquare$} & \ooops{$\blacksquare$} \\
  CrNbTiZr & 2250 &  $-54$ &    $-39$ & $\circledcirc$ & \bingo{$\circledcirc$} & \ooops{$\blacksquare$} & \ooops{$\blacksquare$} & \ooops{$\blacksquare$} & \bingo{$\circledcirc$} & \ooops{$\blacksquare$} & \ooops{$\blacksquare$} \\
   CrTaTiV & 2398 &  $-45$ &    $-94$ & $\circledcirc$ & \ooops{$\blacksquare$} & \ooops{$\blacksquare$} & \ooops{$\blacksquare$} & \ooops{$\blacksquare$} & \ooops{$\blacksquare$} & \ooops{$\blacksquare$} & \ooops{$\blacksquare$} \\
    CrTaVW & 2837 &  $-44$ &   $-129$ & $\circledcirc$ & \bingo{$\circledcirc$} & \ooops{$\blacksquare$} & \ooops{$\blacksquare$} & \ooops{$\blacksquare$} & \ooops{$\blacksquare$} & \ooops{$\blacksquare$} & \bingo{$\circledcirc$} \\
  HfNbTaTi & 2622 &   $27$ &     $24$ & $\blacksquare$ & \bingo{$\blacksquare$} & \bingo{$\blacksquare$} & \bingo{$\blacksquare$} & \bingo{$\blacksquare$} & \ooops{$\circledcirc$} & \bingo{$\blacksquare$} & \bingo{$\blacksquare$} \\
  HfNbTaZr & 2668 &   $34$ &     $26$ & $\blacksquare$ & \bingo{$\blacksquare$} & \bingo{$\blacksquare$} & \ooops{$\circledcirc$} & \ooops{$\circledcirc$} & \bingo{$\blacksquare$} & \bingo{$\blacksquare$} & \bingo{$\blacksquare$} \\
  HfNbTiZr & 2331 &   $25$ &      $9$ & $\blacksquare$ & \bingo{$\blacksquare$} & \bingo{$\blacksquare$} & \bingo{$\blacksquare$} & \bingo{$\blacksquare$} & \ooops{$\circledcirc$} & \bingo{$\blacksquare$} & \bingo{$\blacksquare$} \\
  HfScTiZr & 2097 &   $43$ &     $-8$ & $\circledcirc$ & \ooops{$\blacksquare$} & \ooops{$\blacksquare$} & \bingo{$\circledcirc$} & \bingo{$\circledcirc$} & \bingo{$\circledcirc$} & \ooops{$\blacksquare$} & \ooops{$\blacksquare$} \\
  HfTaTiZr & 2466 &   $18$ &     $25$ & $\blacksquare$ & \bingo{$\blacksquare$} & \bingo{$\blacksquare$} & \bingo{$\blacksquare$} & \bingo{$\blacksquare$} & \ooops{$\circledcirc$} & \bingo{$\blacksquare$} & \bingo{$\blacksquare$} \\
   HfTiYZr & 2094 &   $92$ &     $48$ & $\circledcirc$ & \bingo{$\circledcirc$} & \bingo{$\circledcirc$} & \bingo{$\circledcirc$} & \bingo{$\circledcirc$} & \bingo{$\circledcirc$} & \ooops{$\blacksquare$} & \ooops{$\blacksquare$} \\
  MoNbTaTi & 2719 &  $-28$ &   $-113$ & $\blacksquare$ & \bingo{$\blacksquare$} & \bingo{$\blacksquare$} & \bingo{$\blacksquare$} & \bingo{$\blacksquare$} & \bingo{$\blacksquare$} & \ooops{$\circledcirc$} & \ooops{$\circledcirc$} \\
   MoNbTaV & 2780 &  $-33$ &   $-156$ & $\blacksquare$ & \bingo{$\blacksquare$} & \bingo{$\blacksquare$} & \bingo{$\blacksquare$} & \bingo{$\blacksquare$} & \bingo{$\blacksquare$} & \ooops{$\circledcirc$} & \ooops{$\circledcirc$} \\
   MoNbTaW & 3158 &  $-69$ &   $-131$ & $\blacksquare$ & \bingo{$\blacksquare$} & \bingo{$\blacksquare$} & \bingo{$\blacksquare$} & \bingo{$\blacksquare$} & \bingo{$\blacksquare$} & \ooops{$\circledcirc$} & \bingo{$\blacksquare$} \\
   MoNbTiV & 2442 &  $-26$ &   $-106$ & $\blacksquare$ & \bingo{$\blacksquare$} & \bingo{$\blacksquare$} & \bingo{$\blacksquare$} & \bingo{$\blacksquare$} & \bingo{$\blacksquare$} & \bingo{$\blacksquare$} & \ooops{$\circledcirc$} \\
  MoNbTiZr & 2429 &  $-25$ &    $-93$ & $\blacksquare$ & \bingo{$\blacksquare$} & \bingo{$\blacksquare$} & \bingo{$\blacksquare$} & \bingo{$\blacksquare$} & \ooops{$\circledcirc$} & \bingo{$\blacksquare$} & \ooops{$\circledcirc$} \\
  MoPdRhRu & 2392 &  $-90$ &    $-99$ & $\blacksquare$ & \ooops{$\circledcirc$} & \bingo{$\blacksquare$} & \bingo{$\blacksquare$} & \bingo{$\blacksquare$} & \ooops{$\circledcirc$} & \ooops{$\circledcirc$} & \bingo{$\blacksquare$} \\
   MoTaTiV & 2578 &  $-25$ &   $-135$ & $\blacksquare$ & \bingo{$\blacksquare$} & \bingo{$\blacksquare$} & \bingo{$\blacksquare$} & \bingo{$\blacksquare$} & \bingo{$\blacksquare$} & \ooops{$\circledcirc$} & \ooops{$\circledcirc$} \\
   NbTaTiV & 2541 &   $-1$ &    $-25$ & $\blacksquare$ & \bingo{$\blacksquare$} & \bingo{$\blacksquare$} & \bingo{$\blacksquare$} & \bingo{$\blacksquare$} & \bingo{$\blacksquare$} & \bingo{$\blacksquare$} & \bingo{$\blacksquare$} \\
   NbTaTiW & 2919 &  $-47$ &    $-58$ & $\blacksquare$ & \bingo{$\blacksquare$} & \bingo{$\blacksquare$} & \bingo{$\blacksquare$} & \bingo{$\blacksquare$} & \bingo{$\blacksquare$} & \bingo{$\blacksquare$} & \bingo{$\blacksquare$} \\
  NbTaTiZr & 2527 &   $25$ &     $28$ & $\blacksquare$ & \bingo{$\blacksquare$} & \bingo{$\blacksquare$} & \bingo{$\blacksquare$} & \bingo{$\blacksquare$} & \ooops{$\circledcirc$} & \bingo{$\blacksquare$} & \bingo{$\blacksquare$} \\
    NbTaVW & 2980 &  $-48$ &   $-112$ & $\blacksquare$ & \bingo{$\blacksquare$} & \bingo{$\blacksquare$} & \bingo{$\blacksquare$} & \bingo{$\blacksquare$} & \ooops{$\circledcirc$} & \bingo{$\blacksquare$} & \bingo{$\blacksquare$} \\
   NbTiVZr & 2250 &   $-2$ &     $16$ & $\blacksquare$ & \bingo{$\blacksquare$} & \bingo{$\blacksquare$} & \bingo{$\blacksquare$} & \bingo{$\blacksquare$} & \ooops{$\circledcirc$} & \bingo{$\blacksquare$} & \bingo{$\blacksquare$} \\
  NiPdPtRh & 1959 &   $-9$ &    $-34$ & $\blacksquare$ & \bingo{$\blacksquare$} & \bingo{$\blacksquare$} & \bingo{$\blacksquare$} & \bingo{$\blacksquare$} & \bingo{$\blacksquare$} & \bingo{$\blacksquare$} & \bingo{$\blacksquare$} \\
AgAuCuPdPt & 1560 &  $-65$ &    $-99$ & $\blacksquare$ & \bingo{$\blacksquare$} & \ooops{$\circledcirc$} & \ooops{$\circledcirc$} & \ooops{$\circledcirc$} & \ooops{$\circledcirc$} & \bingo{$\blacksquare$} & \bingo{$\blacksquare$} \\
AlCoCrCuFe & 1610 &  $-37$ &   $-169$ & $\circledcirc$ & \bingo{$\circledcirc$} & \ooops{$\blacksquare$} & \ooops{$\blacksquare$} & \ooops{$\blacksquare$} & \bingo{$\circledcirc$} & \bingo{$\circledcirc$} & \bingo{$\circledcirc$} \\
AlCoCrFeNi & 1684 & $-128$ &   $-300$ & $\blacksquare$ & \ooops{$\circledcirc$} & \bingo{$\blacksquare$} & \ooops{$\circledcirc$} & \ooops{$\circledcirc$} & \ooops{$\circledcirc$} & \ooops{$\circledcirc$} & \ooops{$\circledcirc$} \\
AlCoCuFeNi & 1520 &  $-65$ &   $-294$ & $\circledcirc$ & \bingo{$\circledcirc$} & \ooops{$\blacksquare$} & \ooops{$\blacksquare$} & \ooops{$\blacksquare$} & \ooops{$\blacksquare$} & \bingo{$\circledcirc$} & \bingo{$\circledcirc$} \\
AlCoCuNiZn & 1296 &  $-93$ &   $-294$ & $\circledcirc$ & \bingo{$\circledcirc$} & \ooops{$\blacksquare$} & \ooops{$\blacksquare$} & \ooops{$\blacksquare$} & \ooops{$\blacksquare$} & \bingo{$\circledcirc$} & \bingo{$\circledcirc$} \\
AlCoFeNiTi & 1636 & $-271$ &   $-557$ & $\blacksquare$ & \ooops{$\circledcirc$} & \ooops{$\circledcirc$} & \ooops{$\circledcirc$} & \ooops{$\circledcirc$} & \ooops{$\circledcirc$} & \ooops{$\circledcirc$} & \ooops{$\circledcirc$} \\
AlCrCuFeMg & 1441 &   $50$ &    $-57$ & $\circledcirc$ & \bingo{$\circledcirc$} & \ooops{$\blacksquare$} & \bingo{$\circledcirc$} & \bingo{$\circledcirc$} & \bingo{$\circledcirc$} & \bingo{$\circledcirc$} & \ooops{$\blacksquare$} \\
AlCrCuFeMn & 1560 &  $-29$ &   $-119$ & $\circledcirc$ & \bingo{$\circledcirc$} & \ooops{$\blacksquare$} & \ooops{$\blacksquare$} & \ooops{$\blacksquare$} & \bingo{$\circledcirc$} & \bingo{$\circledcirc$} & \bingo{$\circledcirc$} \\
AlCrCuFeNi & 1602 &  $-53$ &   $-194$ & $\circledcirc$ & \bingo{$\circledcirc$} & \ooops{$\blacksquare$} & \ooops{$\blacksquare$} & \ooops{$\blacksquare$} & \bingo{$\circledcirc$} & \bingo{$\circledcirc$} & \bingo{$\circledcirc$} \\
AlCrFeMoNi & 1910 & $-111$ &   $-251$ & $\circledcirc$ & \bingo{$\circledcirc$} & \ooops{$\blacksquare$} & \ooops{$\blacksquare$} & \ooops{$\blacksquare$} & \bingo{$\circledcirc$} & \bingo{$\circledcirc$} & \bingo{$\circledcirc$} \\
AlCrMoNbTi & 2140 & $-140$ &   $-279$ & $\circledcirc$ & \bingo{$\circledcirc$} & \ooops{$\blacksquare$} & \bingo{$\circledcirc$} & \bingo{$\circledcirc$} & \bingo{$\circledcirc$} & \bingo{$\circledcirc$} & \bingo{$\circledcirc$} \\
AlCrMoSiTi & 1927 & $-241$ &   $-419$ & $\circledcirc$ & \bingo{$\circledcirc$} & \bingo{$\circledcirc$} & \bingo{$\circledcirc$} & \bingo{$\circledcirc$} & \bingo{$\circledcirc$} & \bingo{$\circledcirc$} & \bingo{$\circledcirc$} \\
 AlCrMoTiW & 2329 & $-103$ &   $-223$ & $\blacksquare$ & \bingo{$\blacksquare$} & \bingo{$\blacksquare$} & \bingo{$\blacksquare$} & \bingo{$\blacksquare$} & \ooops{$\circledcirc$} & \ooops{$\circledcirc$} & \ooops{$\circledcirc$} \\
 AlCrNbTiV & 1997 & $-151$ &   $-249$ & $\circledcirc$ & \ooops{$\blacksquare$} & \ooops{$\blacksquare$} & \bingo{$\circledcirc$} & \bingo{$\circledcirc$} & \bingo{$\circledcirc$} & \bingo{$\circledcirc$} & \bingo{$\circledcirc$} \\
AlCuFeNiTi & 1554 & $-192$ &   $-425$ & $\blacksquare$ & \ooops{$\circledcirc$} & \bingo{$\blacksquare$} & \ooops{$\circledcirc$} & \ooops{$\circledcirc$} & \ooops{$\circledcirc$} & \ooops{$\circledcirc$} & \ooops{$\circledcirc$} \\
AlCuMnNiPt & 1516 & $-231$ &   $-479$ & $\blacksquare$ & \ooops{$\circledcirc$} & \ooops{$\circledcirc$} & \ooops{$\circledcirc$} & \ooops{$\circledcirc$} & \bingo{$\blacksquare$} & \ooops{$\circledcirc$} & \ooops{$\circledcirc$} \\
  AlCuTaVW & 2292 &  $-41$ &   $-181$ & $\circledcirc$ & \bingo{$\circledcirc$} & \ooops{$\blacksquare$} & \ooops{$\blacksquare$} & \ooops{$\blacksquare$} & \bingo{$\circledcirc$} & \bingo{$\circledcirc$} & \bingo{$\circledcirc$} \\
 AlCuTiYZr & 1632 & $-245$ &   $-339$ & $\circledcirc$ & \bingo{$\circledcirc$} & \bingo{$\circledcirc$} & \bingo{$\circledcirc$} & \bingo{$\circledcirc$} & \bingo{$\circledcirc$} & \bingo{$\circledcirc$} & \bingo{$\circledcirc$} \\
 AlMoNbTiV & 2141 & $-131$ &   $-306$ & $\blacksquare$ & \bingo{$\blacksquare$} & \bingo{$\blacksquare$} & \ooops{$\circledcirc$} & \ooops{$\circledcirc$} & \ooops{$\circledcirc$} & \ooops{$\circledcirc$} & \ooops{$\circledcirc$} \\
 AlMoTaTiV & 2249 & $-132$ &   $-307$ & $\blacksquare$ & \bingo{$\blacksquare$} & \bingo{$\blacksquare$} & \ooops{$\circledcirc$} & \ooops{$\circledcirc$} & \ooops{$\circledcirc$} & \ooops{$\circledcirc$} & \ooops{$\circledcirc$} \\
 AlNbTaTiV & 2219 & $-138$ &   $-251$ & $\blacksquare$ & \bingo{$\blacksquare$} & \bingo{$\blacksquare$} & \ooops{$\circledcirc$} & \ooops{$\circledcirc$} & \ooops{$\circledcirc$} & \ooops{$\circledcirc$} & \ooops{$\circledcirc$} \\
AlNbTaTiZr & 2208 & $-167$ &   $-257$ & $\circledcirc$ & \ooops{$\blacksquare$} & \bingo{$\circledcirc$} & \bingo{$\circledcirc$} & \bingo{$\circledcirc$} & \bingo{$\circledcirc$} & \bingo{$\circledcirc$} & \bingo{$\circledcirc$} \\
 AlNbTiVZr & 1987 & $-180$ &   $-260$ & $\circledcirc$ & \ooops{$\blacksquare$} & \bingo{$\circledcirc$} & \bingo{$\circledcirc$} & \bingo{$\circledcirc$} & \bingo{$\circledcirc$} & \bingo{$\circledcirc$} & \bingo{$\circledcirc$} \\
AuCuPdPtSn & 1414 & $-162$ &   $-309$ & $\blacksquare$ & \ooops{$\circledcirc$} & \bingo{$\blacksquare$} & \ooops{$\circledcirc$} & \ooops{$\circledcirc$} & \bingo{$\blacksquare$} & \ooops{$\circledcirc$} & \ooops{$\circledcirc$} \\
CoCrCuFeNi & 1769 &   $34$ &      $8$ & $\circledcirc$ & \bingo{$\circledcirc$} & \ooops{$\blacksquare$} & \bingo{$\circledcirc$} & \bingo{$\circledcirc$} & \ooops{$\blacksquare$} & \ooops{$\blacksquare$} & \ooops{$\blacksquare$} \\
CoCrCuNiZn & 1545 &   $-3$ &    $-44$ & $\blacksquare$ & \ooops{$\circledcirc$} & \bingo{$\blacksquare$} & \bingo{$\blacksquare$} & \bingo{$\blacksquare$} & \bingo{$\blacksquare$} & \ooops{$\circledcirc$} & \bingo{$\blacksquare$} \\
CoCrFeHfNi & 1999 & $-202$ &   $-254$ & $\circledcirc$ & \bingo{$\circledcirc$} & \ooops{$\blacksquare$} & \bingo{$\circledcirc$} & \bingo{$\circledcirc$} & \bingo{$\circledcirc$} & \bingo{$\circledcirc$} & \ooops{$\blacksquare$} \\
CoCrFeMnNi & 1801 &  $-43$ &    $-64$ & $\blacksquare$ & \bingo{$\blacksquare$} & \bingo{$\blacksquare$} & \bingo{$\blacksquare$} & \bingo{$\blacksquare$} & \bingo{$\blacksquare$} & \bingo{$\blacksquare$} & \bingo{$\blacksquare$} \\
CoCrFeMoNi & 2077 &  $-48$ &    $-45$ & $\circledcirc$ & \bingo{$\circledcirc$} & \ooops{$\blacksquare$} & \ooops{$\blacksquare$} & \ooops{$\blacksquare$} & \ooops{$\blacksquare$} & \ooops{$\blacksquare$} & \ooops{$\blacksquare$} \\
CoCrFeNbNi & 2047 & $-153$ &   $-131$ & $\circledcirc$ & \bingo{$\circledcirc$} & \ooops{$\blacksquare$} & \bingo{$\circledcirc$} & \bingo{$\circledcirc$} & \ooops{$\blacksquare$} & \bingo{$\circledcirc$} & \ooops{$\blacksquare$} \\
CoCrFeNiPd & 1863 &  $-59$ &    $-62$ & $\blacksquare$ & \bingo{$\blacksquare$} & \bingo{$\blacksquare$} & \bingo{$\blacksquare$} & \bingo{$\blacksquare$} & \bingo{$\blacksquare$} & \bingo{$\blacksquare$} & \bingo{$\blacksquare$} \\
CoCrFeNiTa & 2155 & $-149$ &   $-180$ & $\circledcirc$ & \bingo{$\circledcirc$} & \ooops{$\blacksquare$} & \bingo{$\circledcirc$} & \bingo{$\circledcirc$} & \ooops{$\blacksquare$} & \bingo{$\circledcirc$} & \ooops{$\blacksquare$} \\
CoCrFeNiTi & 1886 & $-169$ &   $-258$ & $\blacksquare$ & \ooops{$\circledcirc$} & \bingo{$\blacksquare$} & \ooops{$\circledcirc$} & \ooops{$\circledcirc$} & \bingo{$\blacksquare$} & \ooops{$\circledcirc$} & \ooops{$\circledcirc$} \\
 CoCrFeNiV & 1934 &  $-93$ &   $-139$ & $\circledcirc$ & \ooops{$\blacksquare$} & \ooops{$\blacksquare$} & \ooops{$\blacksquare$} & \ooops{$\blacksquare$} & \ooops{$\blacksquare$} & \bingo{$\circledcirc$} & \ooops{$\blacksquare$} \\
 CoCrFeNiW & 2236 &  $-31$ &    $-59$ & $\circledcirc$ & \bingo{$\circledcirc$} & \ooops{$\blacksquare$} & \ooops{$\blacksquare$} & \ooops{$\blacksquare$} & \ooops{$\blacksquare$} & \ooops{$\blacksquare$} & \ooops{$\blacksquare$} \\
 CoCrFeNiY & 1857 &  $-95$ &   $-116$ & $\circledcirc$ & \bingo{$\circledcirc$} & \bingo{$\circledcirc$} & \bingo{$\circledcirc$} & \bingo{$\circledcirc$} & \bingo{$\circledcirc$} & \bingo{$\circledcirc$} & \ooops{$\blacksquare$} \\
CoCrFeNiZr & 1923 & $-233$ &   $-206$ & $\circledcirc$ & \bingo{$\circledcirc$} & \ooops{$\blacksquare$} & \bingo{$\circledcirc$} & \bingo{$\circledcirc$} & \bingo{$\circledcirc$} & \bingo{$\circledcirc$} & \ooops{$\blacksquare$} \\
 CoCrMnNiV & 1876 &  $-95$ &   $-170$ & $\circledcirc$ & \ooops{$\blacksquare$} & \ooops{$\blacksquare$} & \ooops{$\blacksquare$} & \ooops{$\blacksquare$} & \ooops{$\blacksquare$} & \bingo{$\circledcirc$} & \bingo{$\circledcirc$} \\
CoCrMoNbTi & 2307 & $-139$ &   $-160$ & $\circledcirc$ & \bingo{$\circledcirc$} & \ooops{$\blacksquare$} & \bingo{$\circledcirc$} & \bingo{$\circledcirc$} & \ooops{$\blacksquare$} & \bingo{$\circledcirc$} & \ooops{$\blacksquare$} \\
 CoCrNiTiV & 1960 & $-195$ &   $-245$ & $\circledcirc$ & \bingo{$\circledcirc$} & \ooops{$\blacksquare$} & \bingo{$\circledcirc$} & \bingo{$\circledcirc$} & \ooops{$\blacksquare$} & \bingo{$\circledcirc$} & \ooops{$\blacksquare$} \\
CoCuFeMnNi & 1637 &   $18$ &    $-42$ & $\blacksquare$ & \ooops{$\circledcirc$} & \bingo{$\blacksquare$} & \bingo{$\blacksquare$} & \bingo{$\blacksquare$} & \bingo{$\blacksquare$} & \bingo{$\blacksquare$} & \bingo{$\blacksquare$} \\
CoCuFeMoNi & 1912 &   $41$ &    $-21$ & $\blacksquare$ & \ooops{$\circledcirc$} & \bingo{$\blacksquare$} & \ooops{$\circledcirc$} & \ooops{$\circledcirc$} & \bingo{$\blacksquare$} & \bingo{$\blacksquare$} & \bingo{$\blacksquare$} \\
CoCuFeNiPd & 1699 &    $2$ &    $-50$ & $\blacksquare$ & \ooops{$\circledcirc$} & \bingo{$\blacksquare$} & \bingo{$\blacksquare$} & \bingo{$\blacksquare$} & \bingo{$\blacksquare$} & \bingo{$\blacksquare$} & \bingo{$\blacksquare$} \\
CoCuFeNiPt & 1741 &  $-27$ &   $-112$ & $\blacksquare$ & \bingo{$\blacksquare$} & \bingo{$\blacksquare$} & \bingo{$\blacksquare$} & \bingo{$\blacksquare$} & \bingo{$\blacksquare$} & \ooops{$\circledcirc$} & \ooops{$\circledcirc$} \\
CoCuFeNiRu & 1854 &   $38$ &     $21$ & $\blacksquare$ & \bingo{$\blacksquare$} & \bingo{$\blacksquare$} & \ooops{$\circledcirc$} & \ooops{$\circledcirc$} & \bingo{$\blacksquare$} & \bingo{$\blacksquare$} & \bingo{$\blacksquare$} \\
CoCuFeNiTi & 1721 & $-113$ &   $-243$ & $\blacksquare$ & \ooops{$\circledcirc$} & \bingo{$\blacksquare$} & \bingo{$\blacksquare$} & \bingo{$\blacksquare$} & \bingo{$\blacksquare$} & \ooops{$\circledcirc$} & \ooops{$\circledcirc$} \\
CoCuNiPdPt & 1745 &  $-45$ &    $-88$ & $\blacksquare$ & \ooops{$\circledcirc$} & \bingo{$\blacksquare$} & \bingo{$\blacksquare$} & \bingo{$\blacksquare$} & \bingo{$\blacksquare$} & \bingo{$\blacksquare$} & \bingo{$\blacksquare$} \\
CoFeIrPdPt & 2037 &  $-51$ &    $-99$ & $\blacksquare$ & \bingo{$\blacksquare$} & \bingo{$\blacksquare$} & \bingo{$\blacksquare$} & \bingo{$\blacksquare$} & \bingo{$\blacksquare$} & \ooops{$\circledcirc$} & \bingo{$\blacksquare$} \\
CoFeMnMoNi & 1944 &  $-41$ &   $-103$ & $\circledcirc$ & \bingo{$\circledcirc$} & \ooops{$\blacksquare$} & \ooops{$\blacksquare$} & \ooops{$\blacksquare$} & \ooops{$\blacksquare$} & \ooops{$\blacksquare$} & \ooops{$\blacksquare$} \\
CoFeMnNiTi & 1753 & $-171$ &   $-324$ & $\circledcirc$ & \bingo{$\circledcirc$} & \ooops{$\blacksquare$} & \bingo{$\circledcirc$} & \bingo{$\circledcirc$} & \ooops{$\blacksquare$} & \bingo{$\circledcirc$} & \bingo{$\circledcirc$} \\
 CoFeMnNiV & 1802 &  $-92$ &   $-211$ & $\circledcirc$ & \ooops{$\blacksquare$} & \ooops{$\blacksquare$} & \ooops{$\blacksquare$} & \ooops{$\blacksquare$} & \ooops{$\blacksquare$} & \bingo{$\circledcirc$} & \bingo{$\circledcirc$} \\
 CoFeMoNiV & 2077 &  $-92$ &   $-170$ & $\circledcirc$ & \ooops{$\blacksquare$} & \ooops{$\blacksquare$} & \ooops{$\blacksquare$} & \ooops{$\blacksquare$} & \ooops{$\blacksquare$} & \bingo{$\circledcirc$} & \bingo{$\circledcirc$} \\
CoFeNiPdPt & 1835 &  $-52$ &   $-131$ & $\blacksquare$ & \bingo{$\blacksquare$} & \bingo{$\blacksquare$} & \bingo{$\blacksquare$} & \bingo{$\blacksquare$} & \bingo{$\blacksquare$} & \ooops{$\circledcirc$} & \ooops{$\circledcirc$} \\
CoIrNiRhRu & 2216 &  $-12$ &     $11$ & $\blacksquare$ & \bingo{$\blacksquare$} & \bingo{$\blacksquare$} & \bingo{$\blacksquare$} & \bingo{$\blacksquare$} & \bingo{$\blacksquare$} & \bingo{$\blacksquare$} & \bingo{$\blacksquare$} \\
CrCuFeMnNi & 1719 &   $28$ &     $-9$ & $\circledcirc$ & \bingo{$\circledcirc$} & \ooops{$\blacksquare$} & \ooops{$\blacksquare$} & \ooops{$\blacksquare$} & \ooops{$\blacksquare$} & \ooops{$\blacksquare$} & \ooops{$\blacksquare$} \\
CrCuFeMoNi & 1995 &   $48$ &     $17$ & $\blacksquare$ & \ooops{$\circledcirc$} & \bingo{$\blacksquare$} & \ooops{$\circledcirc$} & \ooops{$\circledcirc$} & \bingo{$\blacksquare$} & \bingo{$\blacksquare$} & \bingo{$\blacksquare$} \\
CrFeMnNiTi & 1836 & $-137$ &   $-255$ & $\circledcirc$ & \bingo{$\circledcirc$} & \ooops{$\blacksquare$} & \bingo{$\circledcirc$} & \bingo{$\circledcirc$} & \ooops{$\blacksquare$} & \bingo{$\circledcirc$} & \bingo{$\circledcirc$} \\
CrFeMoNbTi & 2316 &  $-96$ &   $-150$ & $\circledcirc$ & \bingo{$\circledcirc$} & \ooops{$\blacksquare$} & \ooops{$\blacksquare$} & \ooops{$\blacksquare$} & \ooops{$\blacksquare$} & \bingo{$\circledcirc$} & \ooops{$\blacksquare$} \\
 CrFeMoNbV & 2364 &  $-69$ &   $-108$ & $\circledcirc$ & \bingo{$\circledcirc$} & \ooops{$\blacksquare$} & \ooops{$\blacksquare$} & \ooops{$\blacksquare$} & \ooops{$\blacksquare$} & \ooops{$\blacksquare$} & \ooops{$\blacksquare$} \\
CrHfNbTiZr & 2301 &  $-44$ &    $-46$ & $\circledcirc$ & \bingo{$\circledcirc$} & \ooops{$\blacksquare$} & \ooops{$\blacksquare$} & \ooops{$\blacksquare$} & \bingo{$\circledcirc$} & \ooops{$\blacksquare$} & \ooops{$\blacksquare$} \\
 CrMoNbTiW & 2692 &  $-58$ &    $-86$ & $\circledcirc$ & \bingo{$\circledcirc$} & \ooops{$\blacksquare$} & \ooops{$\blacksquare$} & \ooops{$\blacksquare$} & \ooops{$\blacksquare$} & \ooops{$\blacksquare$} & \ooops{$\blacksquare$} \\
 CrNbTiVZr & 2236 &  $-49$ &    $-37$ & $\circledcirc$ & \bingo{$\circledcirc$} & \ooops{$\blacksquare$} & \ooops{$\blacksquare$} & \ooops{$\blacksquare$} & \bingo{$\circledcirc$} & \ooops{$\blacksquare$} & \ooops{$\blacksquare$} \\
CuFeMnNiPt & 1691 &  $-79$ &   $-170$ & $\circledcirc$ & \ooops{$\blacksquare$} & \ooops{$\blacksquare$} & \ooops{$\blacksquare$} & \ooops{$\blacksquare$} & \ooops{$\blacksquare$} & \bingo{$\circledcirc$} & \bingo{$\circledcirc$} \\
 FeMoNbTiV & 2316 &  $-86$ &   $-185$ & $\circledcirc$ & \bingo{$\circledcirc$} & \ooops{$\blacksquare$} & \ooops{$\blacksquare$} & \ooops{$\blacksquare$} & \ooops{$\blacksquare$} & \bingo{$\circledcirc$} & \bingo{$\circledcirc$} \\
HfMoNbTaTi & 2677 &  $-13$ &    $-90$ & $\blacksquare$ & \bingo{$\blacksquare$} & \bingo{$\blacksquare$} & \bingo{$\blacksquare$} & \bingo{$\blacksquare$} & \bingo{$\blacksquare$} & \bingo{$\blacksquare$} & \ooops{$\circledcirc$} \\
HfMoNbTaZr & 2714 &  $-13$ &    $-84$ & $\blacksquare$ & \bingo{$\blacksquare$} & \bingo{$\blacksquare$} & \bingo{$\blacksquare$} & \bingo{$\blacksquare$} & \bingo{$\blacksquare$} & \bingo{$\blacksquare$} & \bingo{$\blacksquare$} \\
HfMoNbTiZr & 2444 &  $-16$ &    $-89$ & $\blacksquare$ & \bingo{$\blacksquare$} & \bingo{$\blacksquare$} & \bingo{$\blacksquare$} & \bingo{$\blacksquare$} & \ooops{$\circledcirc$} & \bingo{$\blacksquare$} & \bingo{$\blacksquare$} \\
HfMoTaTiZr & 2552 &  $-20$ &    $-92$ & $\blacksquare$ & \bingo{$\blacksquare$} & \bingo{$\blacksquare$} & \bingo{$\blacksquare$} & \bingo{$\blacksquare$} & \ooops{$\circledcirc$} & \bingo{$\blacksquare$} & \bingo{$\blacksquare$} \\
HfNbTaTiZr & 2523 &   $28$ &     $24$ & $\blacksquare$ & \bingo{$\blacksquare$} & \bingo{$\blacksquare$} & \bingo{$\blacksquare$} & \bingo{$\blacksquare$} & \ooops{$\circledcirc$} & \bingo{$\blacksquare$} & \bingo{$\blacksquare$} \\
 HfNbTiVZr & 2302 &    $2$ &     $10$ & $\circledcirc$ & \bingo{$\circledcirc$} & \ooops{$\blacksquare$} & \ooops{$\blacksquare$} & \ooops{$\blacksquare$} & \bingo{$\circledcirc$} & \ooops{$\blacksquare$} & \ooops{$\blacksquare$} \\
 HfScTiYZr & 2038 &   $88$ &     $31$ & $\circledcirc$ & \bingo{$\circledcirc$} & \bingo{$\circledcirc$} & \bingo{$\circledcirc$} & \bingo{$\circledcirc$} & \bingo{$\circledcirc$} & \ooops{$\blacksquare$} & \ooops{$\blacksquare$} \\
IrOsReRhRu & 2870 &   $-3$ &   $-134$ & $\blacksquare$ & \bingo{$\blacksquare$} & \bingo{$\blacksquare$} & \bingo{$\blacksquare$} & \bingo{$\blacksquare$} & \bingo{$\blacksquare$} & \ooops{$\circledcirc$} & \ooops{$\circledcirc$} \\
IrPdPtRhRu & 2290 &   $25$ &    $-23$ & $\blacksquare$ & \bingo{$\blacksquare$} & \bingo{$\blacksquare$} & \bingo{$\blacksquare$} & \bingo{$\blacksquare$} & \bingo{$\blacksquare$} & \bingo{$\blacksquare$} & \bingo{$\blacksquare$} \\
 MoNbReTaW & 3218 & $-145$ &   $-196$ & $\blacksquare$ & \bingo{$\blacksquare$} & \bingo{$\blacksquare$} & \ooops{$\circledcirc$} & \ooops{$\circledcirc$} & \bingo{$\blacksquare$} & \ooops{$\circledcirc$} & \bingo{$\blacksquare$} \\
 MoNbTaTiV & 2612 &  $-24$ &   $-114$ & $\blacksquare$ & \bingo{$\blacksquare$} & \bingo{$\blacksquare$} & \bingo{$\blacksquare$} & \bingo{$\blacksquare$} & \bingo{$\blacksquare$} & \bingo{$\blacksquare$} & \ooops{$\circledcirc$} \\
 MoNbTaTiW & 2914 &  $-54$ &   $-118$ & $\blacksquare$ & \bingo{$\blacksquare$} & \bingo{$\blacksquare$} & \bingo{$\blacksquare$} & \bingo{$\blacksquare$} & \bingo{$\blacksquare$} & \bingo{$\blacksquare$} & \bingo{$\blacksquare$} \\
  MoNbTaVW & 2963 &  $-49$ &   $-146$ & $\blacksquare$ & \bingo{$\blacksquare$} & \bingo{$\blacksquare$} & \bingo{$\blacksquare$} & \bingo{$\blacksquare$} & \bingo{$\blacksquare$} & \bingo{$\blacksquare$} & \ooops{$\circledcirc$} \\
 MoNbTiVZr & 2380 &  $-27$ &    $-78$ & $\blacksquare$ & \ooops{$\circledcirc$} & \bingo{$\blacksquare$} & \bingo{$\blacksquare$} & \bingo{$\blacksquare$} & \ooops{$\circledcirc$} & \bingo{$\blacksquare$} & \bingo{$\blacksquare$} \\
  MoTaVWZr & 2838 &  $-51$ &   $-143$ & $\circledcirc$ & \bingo{$\circledcirc$} & \ooops{$\blacksquare$} & \ooops{$\blacksquare$} & \ooops{$\blacksquare$} & \bingo{$\circledcirc$} & \ooops{$\blacksquare$} & \bingo{$\circledcirc$} \\
 NbNiTaTiW & 2681 & $-190$ &   $-237$ & $\circledcirc$ & \bingo{$\circledcirc$} & \ooops{$\blacksquare$} & \bingo{$\circledcirc$} & \bingo{$\circledcirc$} & \ooops{$\blacksquare$} & \bingo{$\circledcirc$} & \ooops{$\blacksquare$} \\
 NbReTaTiV & 2725 & $-146$ &   $-253$ & $\blacksquare$ & \bingo{$\blacksquare$} & \bingo{$\blacksquare$} & \ooops{$\circledcirc$} & \ooops{$\circledcirc$} & \bingo{$\blacksquare$} & \ooops{$\circledcirc$} & \ooops{$\circledcirc$} \\
NbSnTaTiZr & 2123 &  $-97$ &   $-183$ & $\circledcirc$ & \bingo{$\circledcirc$} & \ooops{$\blacksquare$} & \ooops{$\blacksquare$} & \ooops{$\blacksquare$} & \ooops{$\blacksquare$} & \bingo{$\circledcirc$} & \bingo{$\circledcirc$} \\
  NbTaTiVW & 2772 &  $-37$ &    $-73$ & $\blacksquare$ & \bingo{$\blacksquare$} & \bingo{$\blacksquare$} & \bingo{$\blacksquare$} & \bingo{$\blacksquare$} & \bingo{$\blacksquare$} & \bingo{$\blacksquare$} & \bingo{$\blacksquare$} \\
\end{longtable*}
\clearpage

\begingroup
\squeezetable
\begin{table}
\caption{\label{tab:structure}Structures of the 73 experimentally confirmed single-phase HEAs predicted by the present and the VEC model.}
\begin{ruledtabular}
\begin{tabular}{llll}
   Single-phase HEA &   Expt.  & Present & VEC \\
\hline
  AlCoFeNi &      BCC &   BCC &   IM \\
  AlCrFeNi &      BCC &   BCC &  BCC \\
  AlCrMoTi &      BCC &   BCC &  BCC \\
  AlCuNiTi &      FCC &   BCC &   IM \\
  AlMoNbTi &      BCC &   BCC &  BCC \\
  AlNbTaTi &      BCC &   BCC &  BCC \\
   AlNbTiV &      BCC &   BCC &  BCC \\
  CoCrCuFe &      FCC &   FCC &  FCC \\
  CoCrFeNi &      FCC &   FCC &  FCC \\
  CoCrMnNi &      FCC &   HCP &   IM \\
  CoCuFeNi &      FCC &   FCC &  FCC \\
  CoFeMnNi &      FCC &   FCC &  FCC \\
  CoFeNiPd &      FCC &   BCC &  FCC \\
   CoFeNiV &      FCC &   BCC &   IM \\
  CoFeReRu &      HCP &   BCC &   IM \\
  CoNiRhRu &      FCC &   HCP &  FCC \\
  CrFeMnNi &      FCC &   FCC &   IM \\
  HfNbTaTi &      BCC &   BCC &  BCC \\
  HfNbTaZr &      BCC &   BCC &  BCC \\
  HfNbTiZr &      BCC &   BCC &  BCC \\
  HfTaTiZr &      BCC &   HCP &  BCC \\
  MoNbTaTi &      BCC &   BCC &  BCC \\
   MoNbTaV &      BCC &   BCC &  BCC \\
   MoNbTaW &      BCC &   BCC &  BCC \\
   MoNbTiV &      BCC &   BCC &  BCC \\
  MoNbTiZr &      BCC &   BCC &  BCC \\
  MoPdRhRu &      HCP &   FCC &  FCC \\
   MoTaTiV &      BCC &   BCC &  BCC \\
   NbTaTiV &      BCC &   BCC &  BCC \\
   NbTaTiW &      BCC &   BCC &  BCC \\
  NbTaTiZr &      BCC &   BCC &  BCC \\
    NbTaVW &      BCC &   BCC &  BCC \\
   NbTiVZr &      BCC &   BCC &  BCC \\
  NiPdPtRh &      FCC &   FCC &  FCC \\
AgAuCuPdPt &      FCC &   FCC &  FCC \\
AlCoCrFeNi &      BCC &   HCP &   IM \\
AlCoFeNiTi &      BCC &   BCC &  BCC \\
 AlCrMoTiW &      BCC &   BCC &  BCC \\
AlCuFeNiTi &      FCC &   BCC &   IM \\
AlCuMnNiPt &      FCC &   HCP &  FCC \\
 AlMoNbTiV &      BCC &   BCC &  BCC \\
 AlMoTaTiV &      BCC &   BCC &  BCC \\
 AlNbTaTiV &      BCC &   BCC &  BCC \\
AuCuPdPtSn &      FCC &   HCP &  FCC \\
CoCrCuNiZn &      FCC &   FCC &  FCC \\
CoCrFeMnNi &      FCC &   FCC &   IM \\
CoCrFeNiPd &      FCC &   FCC &  FCC \\
CoCrFeNiTi &      FCC &   BCC &   IM \\
CoCuFeMnNi &      FCC &   FCC &  FCC \\
CoCuFeMoNi &      FCC &   FCC &  FCC \\
CoCuFeNiPd &      BCC &   FCC &  FCC \\
CoCuFeNiPt &      FCC &   FCC &  FCC \\
CoCuFeNiRu &      HCP &   BCC &  FCC \\
CoCuFeNiTi &      FCC &   BCC &  FCC \\
CoCuNiPdPt &      FCC &   FCC &  FCC \\
CoFeIrPdPt &      FCC &   FCC &  FCC \\
CoFeNiPdPt &      FCC &   FCC &  FCC \\
CoIrNiRhRu &      FCC &   HCP &  FCC \\
CrCuFeMoNi &      FCC &   FCC &  FCC \\
HfMoNbTaTi &      BCC &   BCC &  BCC \\
HfMoNbTaZr &      BCC &   BCC &  BCC \\
HfMoNbTiZr &      BCC &   BCC &  BCC \\
HfMoTaTiZr &      BCC &   BCC &  BCC \\
HfNbTaTiZr &      BCC &   BCC &  BCC \\
IrOsReRhRu &      HCP &   HCP &  FCC \\
IrPdPtRhRu &      FCC &   FCC &  FCC \\
 MoNbReTaW &      BCC &   BCC &  BCC \\
 MoNbTaTiV &      BCC &   BCC &  BCC \\
 MoNbTaTiW &      BCC &   BCC &  BCC \\
  MoNbTaVW &      BCC &   BCC &  BCC \\
 MoNbTiVZr &      BCC &   BCC &  BCC \\
 NbReTaTiV &      BCC &   BCC &  BCC \\
  NbTaTiVW &      BCC &   BCC &  BCC \\
Accuracy (\%) &       &   74  &   78 \\
\end{tabular}
\end{ruledtabular}
\end{table}
\endgroup

\begin{table*}
\caption{\label{tab:summary}Number ($N$) of quinaries that are predicted 
as stable single-phase solid solutions at $T_a=0.9T_m$ and 1350~K from a total of 
\num{658008} candidates.
The percentage (pct) with respect to the numer of candidates is also given.
The refractory compounds refer to the quinaries comprising at least one refractory element.
}
\begin{ruledtabular}
\begin{tabular}{lllllllllll}
{} & Total & FCC & HCP & BCC & Refractory \\
\hline
$T_a=0.9T_m$\\
$N$ &  \num{30201} &   1664 &   5906 &  \num{22631} & \num{23781} \\
pct (\%) &    4.6 &   0.3 &  0.9 &  3.4 &  3.6 \\
$T_a=1350$~K\\
$N$ &  4512 &      306 &     704 &     3502 &    3722  \\
pct (\%) &    0.7 &      0.0 &      0.1 &      0.5 &   0.6 \\
\end{tabular}
\end{ruledtabular}
\end{table*}

\clearpage

\begin{figure*}
\includegraphics{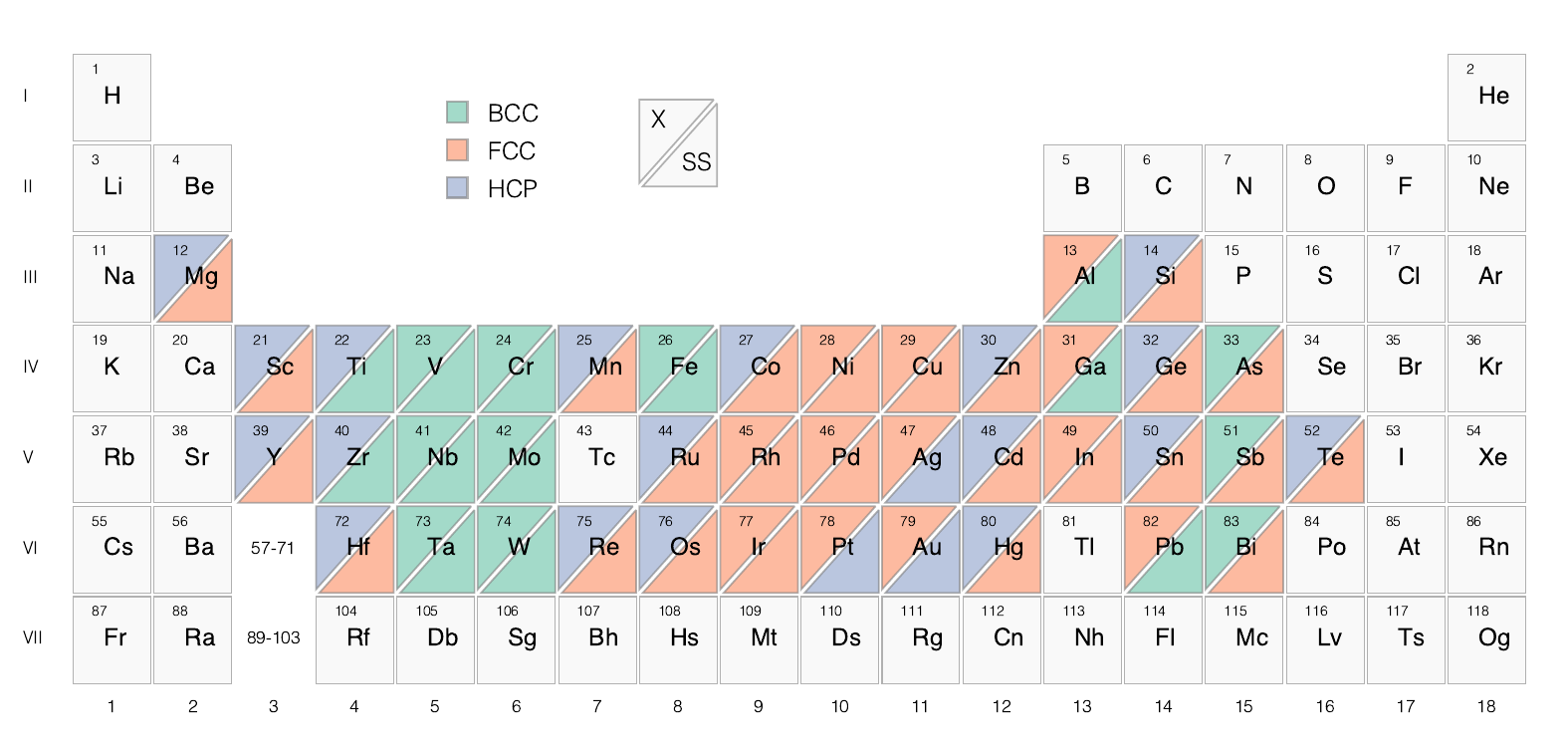}
\caption{\label{fig:pt}Metal elements included in the present study. 
         The ground-state elemental phase at 0 K is shown by the upper triangle, 
         whereas the most stable phase of the element found in its binary solid solutions (SS) is shown by the lower triangle.
}
\end{figure*}

\begin{figure}
\includegraphics{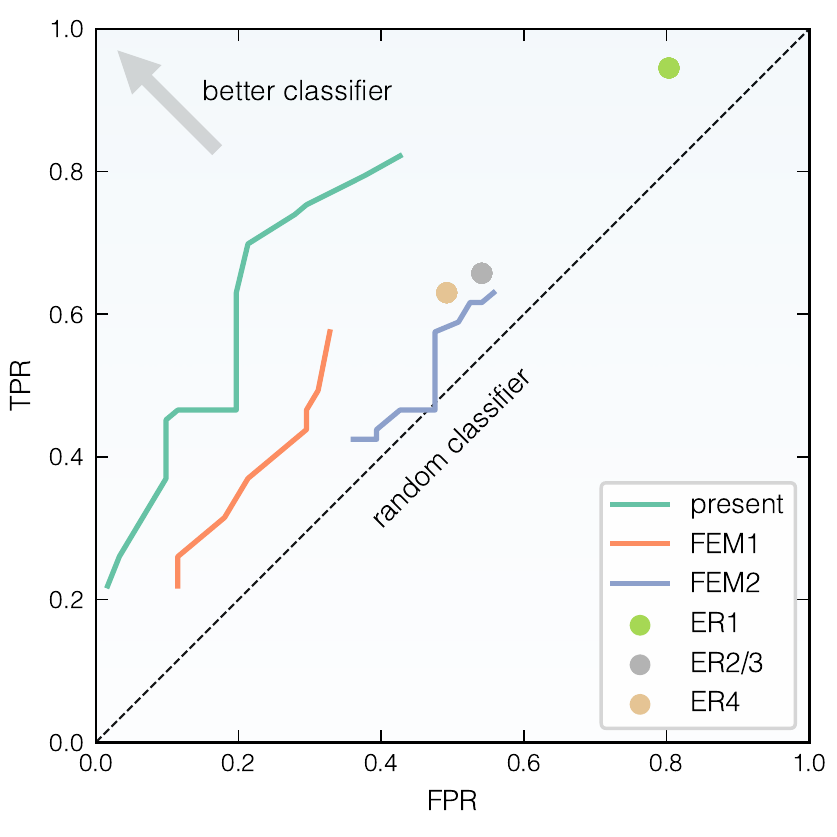}
\caption{\label{fig:roc}True positive ratio (TPR) vs false positive ratio (FPR) for the present and  
         the previously developed models. Except for the ERs, the TRP and FPR are assessed for 
         temperatures ranging from 800 to 1600~K. Analogous to the receiver operating characteristic (ROC) analysis,
         models exhibiting its presence closer to the upper left corner are better at classifying single-phase HEAs and multi-phase alloys.
}
\end{figure}

\begin{figure*}
\includegraphics{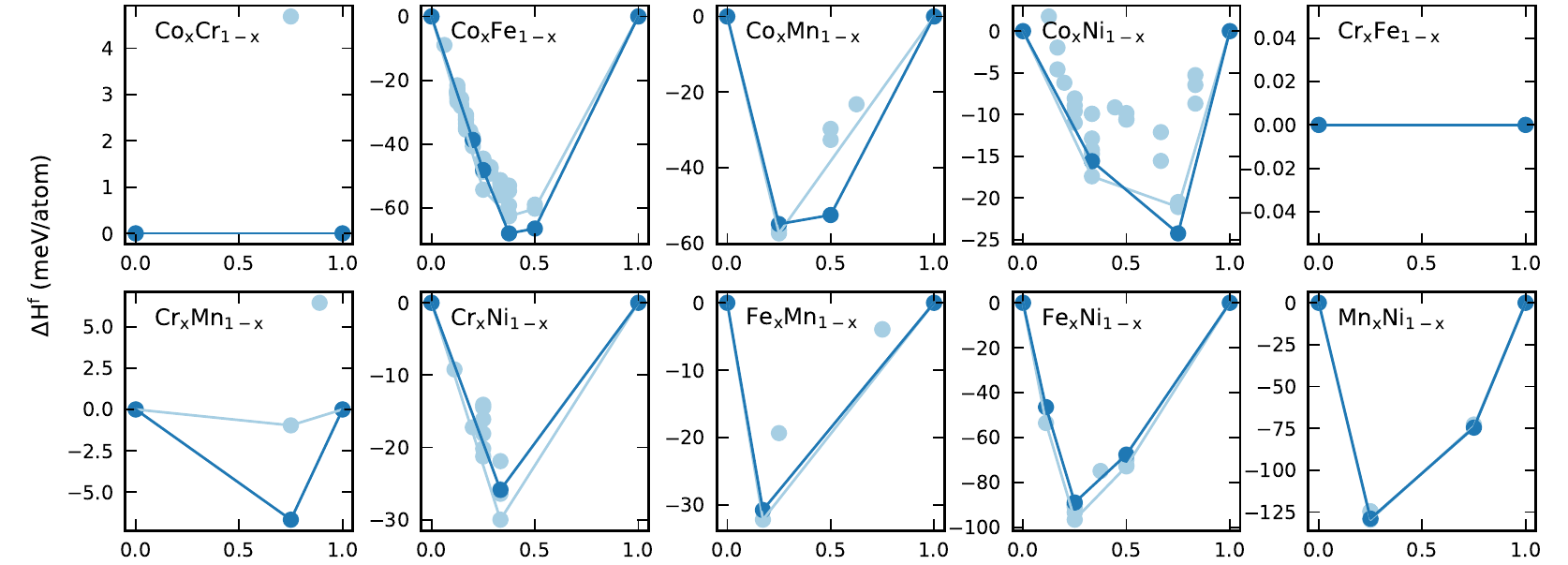}\\
\vspace{1cm}
\includegraphics{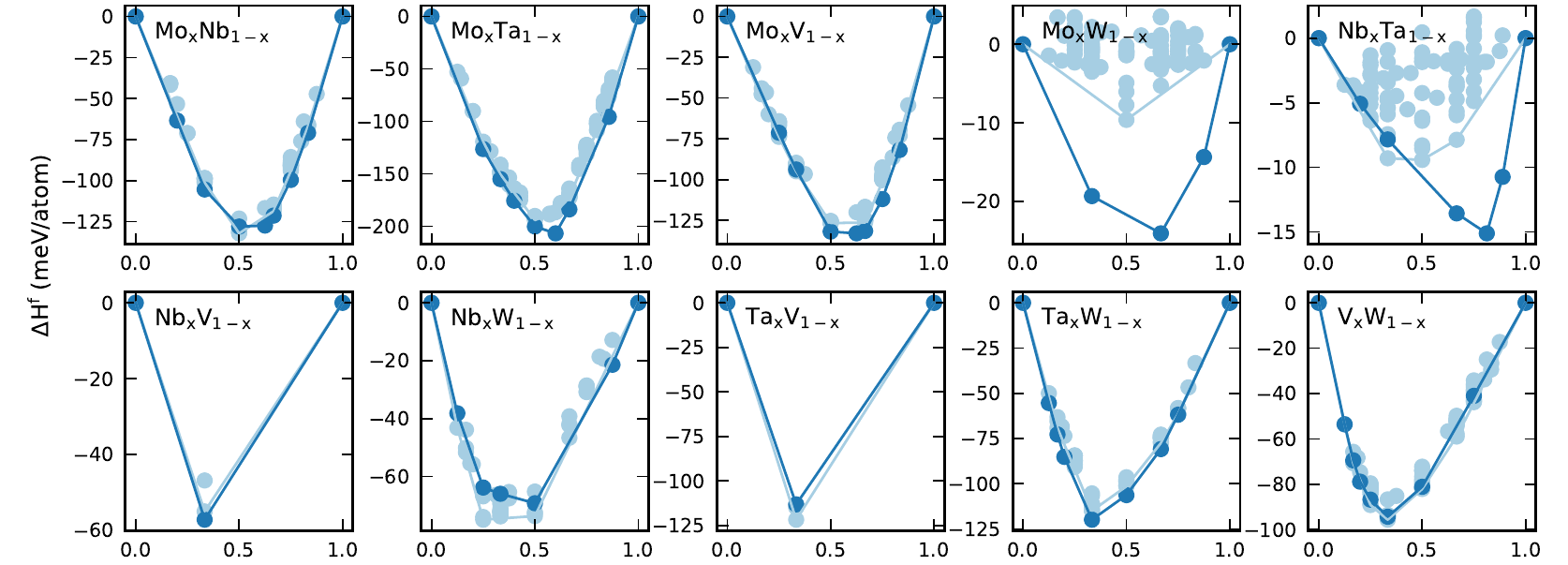}
\caption{\label{fig:pd_comp}Predicted formation enthalpy vs atomic fraction for the binary intermetallic subsystems of the Cantor alloy (\ce{CoCrFeMnNi}) and of the 
refractory alloy (\ce{MoNbTaVW}). The \textsc{Aflow} \textsc{lib2} dataset and the present results are shown by the light and dark blue markers, respectively.
The convex hulls are indicated. For the \textsc{Aflow} dataset an energy above hull cutoff of 10 meV/atom is applied.
The present results contain only the stable compounds.
}
\end{figure*}

\begin{figure*}
\includegraphics{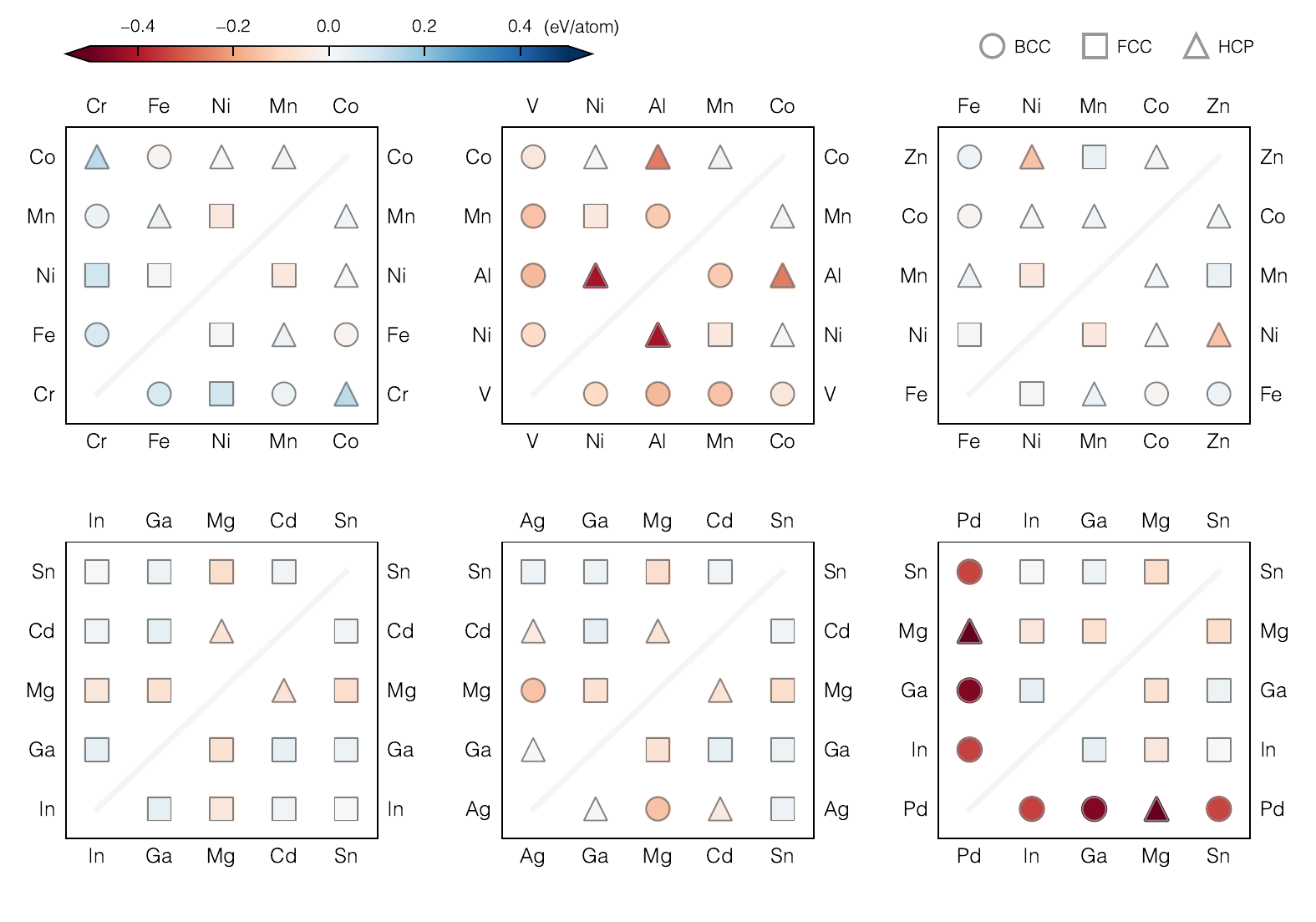}
\caption{\label{fig:submatrix}Predicted formation enthalpy of binary solid solutions for the constituent elements
        of \ce{CoCrFeMnNi}, \ce{AlCoMnNiV}, \ce{CoFeMnNiZn}, \ce{CdGaInMgSn}, \ce{AgCdGaMgSn}, and \ce{GaInMgPdSn}.}
\end{figure*}

\begin{figure*}
\includegraphics{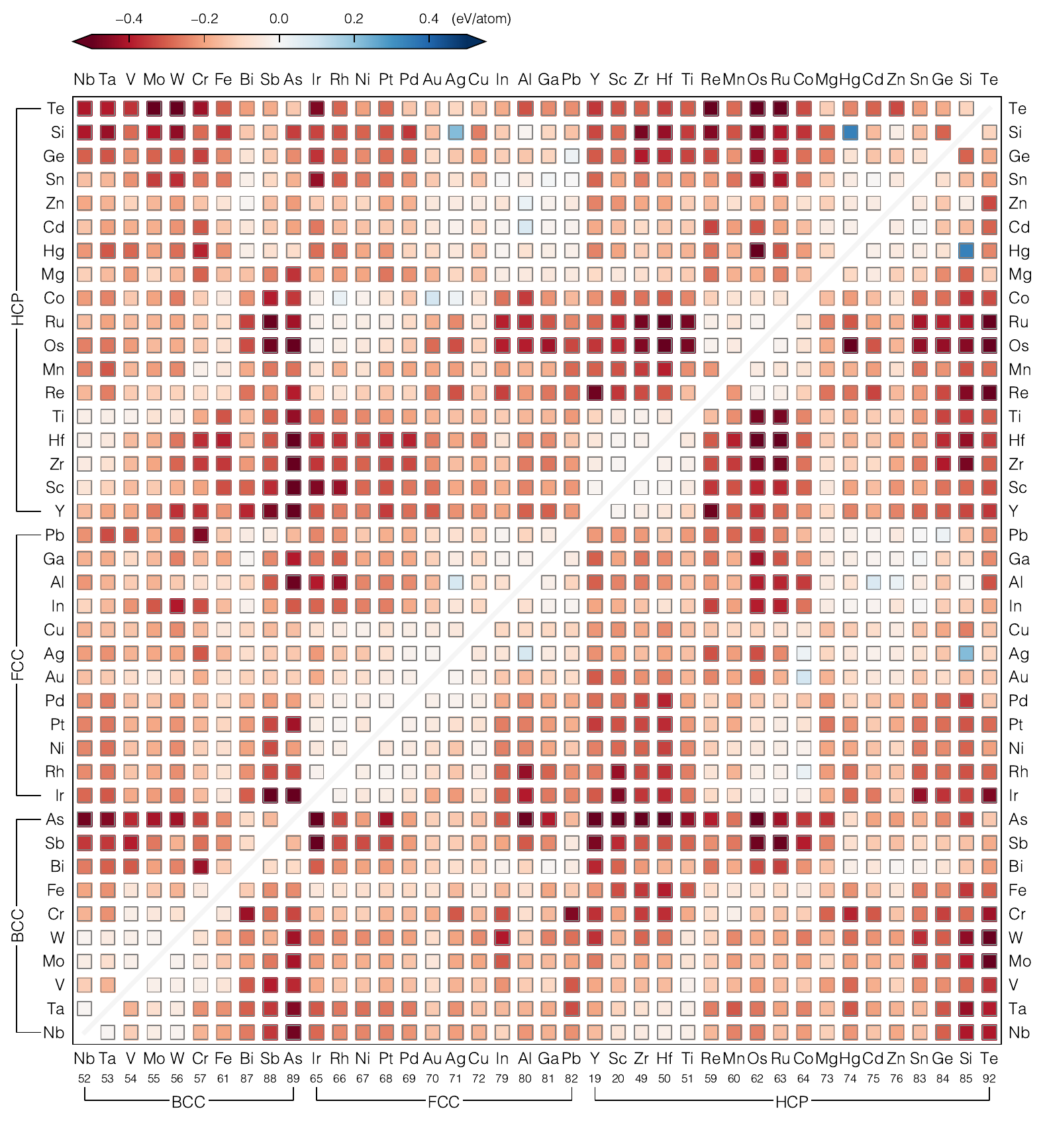}
\caption{\label{fig:im_matrix}Predicted formation enthalpy of the binary intermetallics relative to that of the solid solution. 
        Darker red suggests the presence of stronger competing intermetallic phases for a given binary system.}
\end{figure*}

\begin{figure*}
\includegraphics{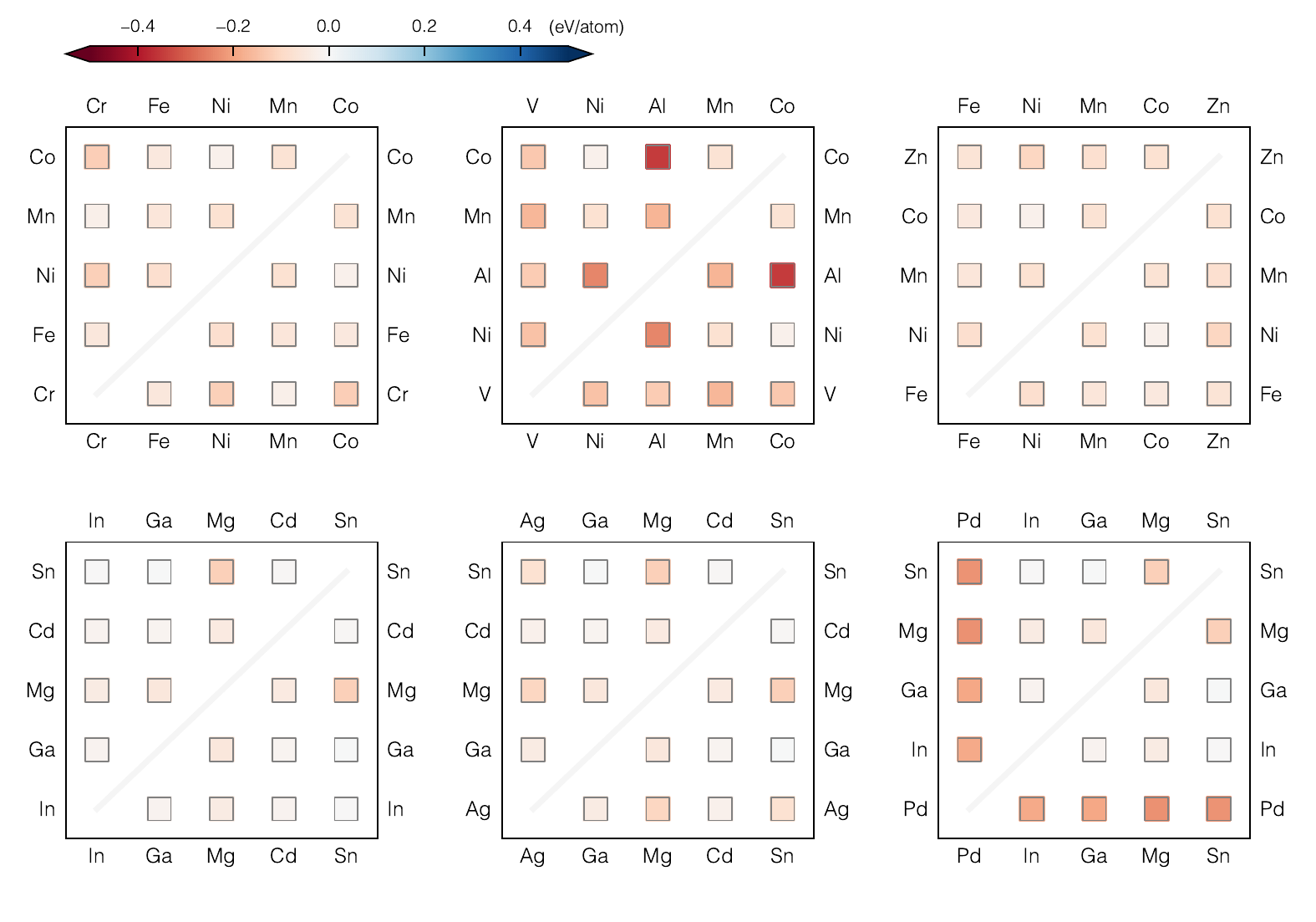}
\caption{\label{fig:im_submatrix}Same as Fig.~\ref{fig:im_matrix} but only for the elements of
        \ce{CoCrFeMnNi}, \ce{AlCoMnNiV}, \ce{CoFeMnNiZn}, \ce{CdGaInMgSn}, \ce{AgCdGaMgSn}, and \ce{GaInMgPdSn}.}
\end{figure*}

\begin{figure*}
\includegraphics{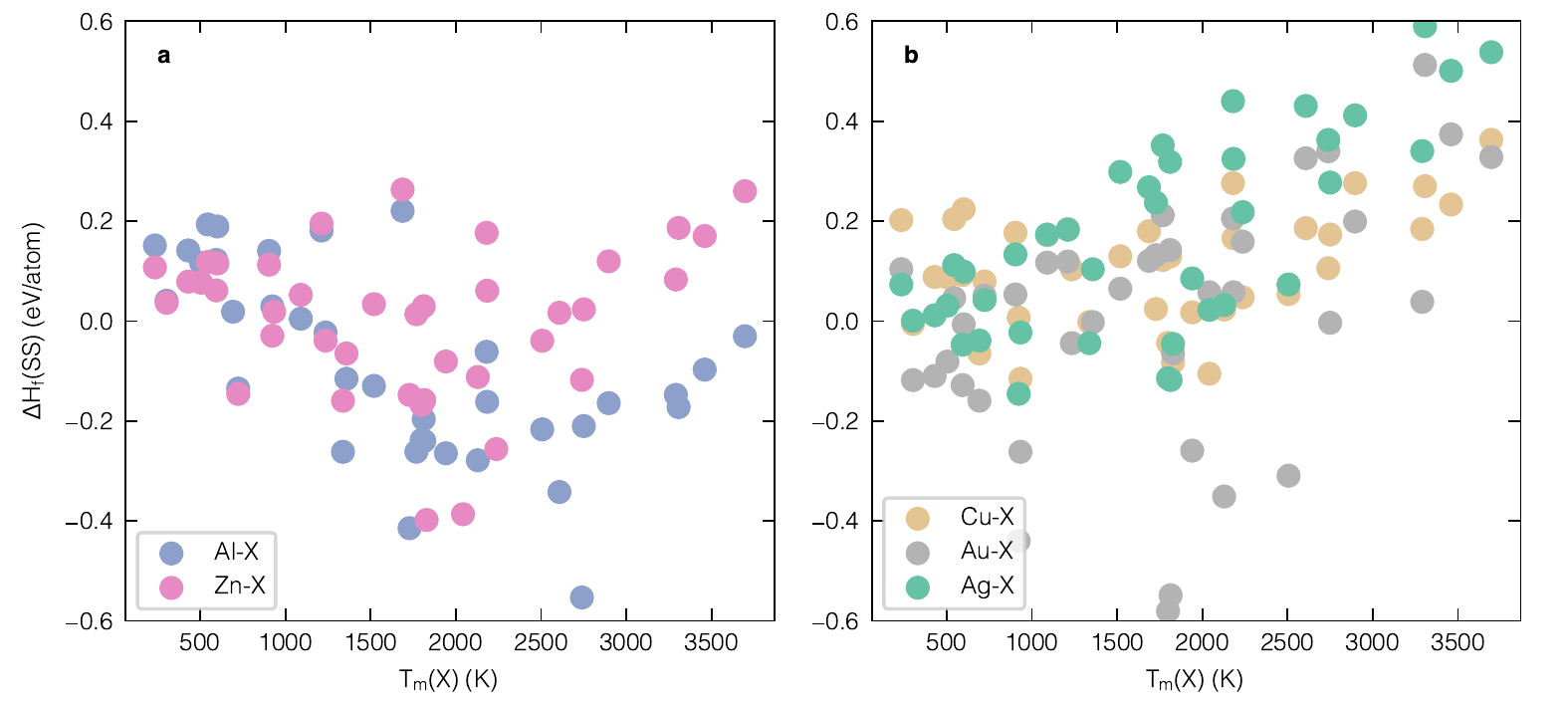}
\caption{\label{fig:form_tm}\textbf{a} Predicted formation enthalpy for Al--$X$ and Zn--$X$ solid solutions as a function of the melting point of element $X$. 
        \textbf{b} is for Cu--$X$, Au--$X$, and Ag--$X$.}
\end{figure*}

\begin{figure*}
\includegraphics{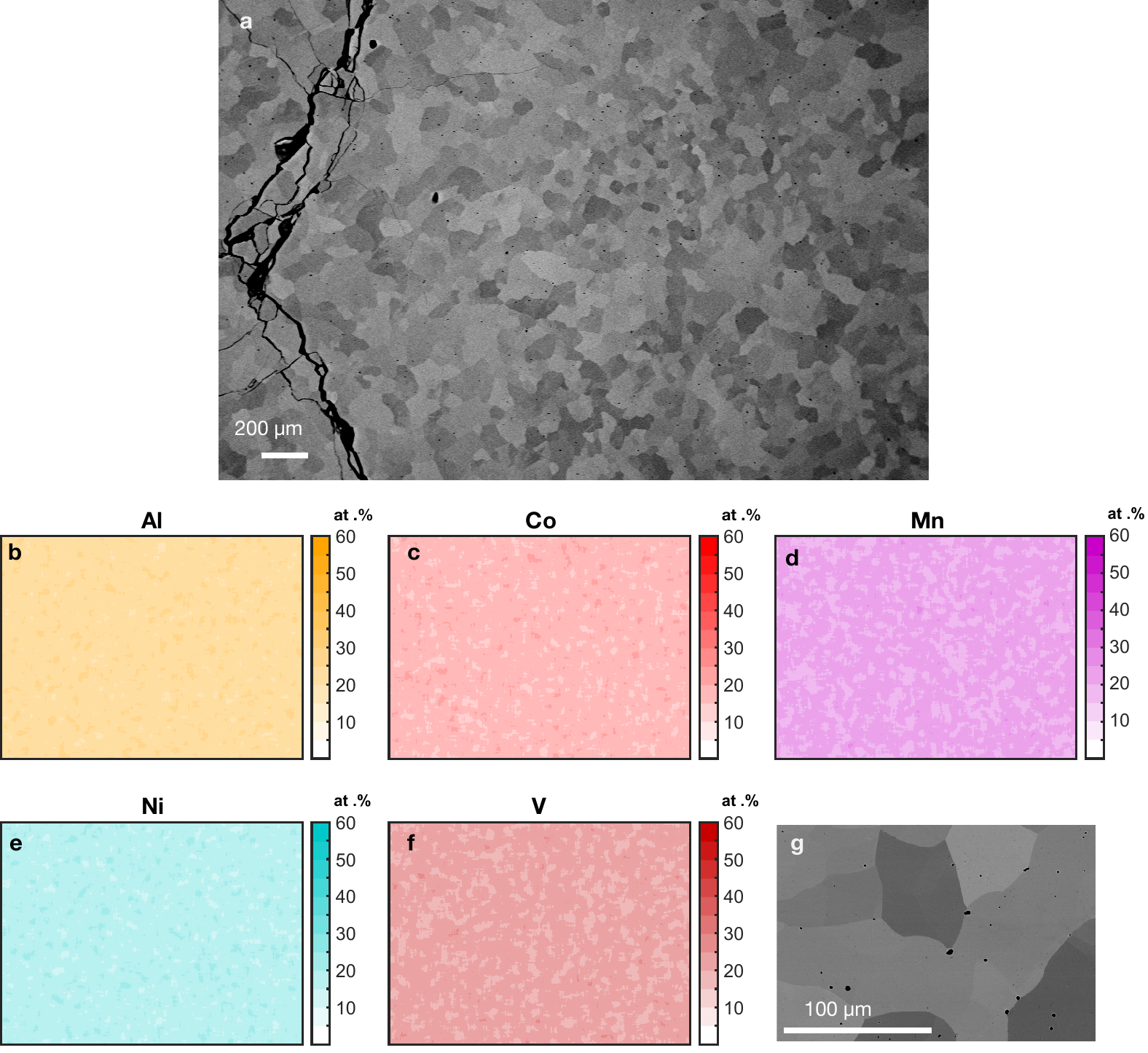}
\caption{\label{fig:supBCC} \textbf{a} Low magnification SEM micrograph of \ce{AlCoMnNiV} taken under secondary-electrons (SE) mode. \textbf{b-f} EDX mapping of the BCC matrix of the region \textbf{g}.}
\end{figure*}

\begin{figure*}
\includegraphics{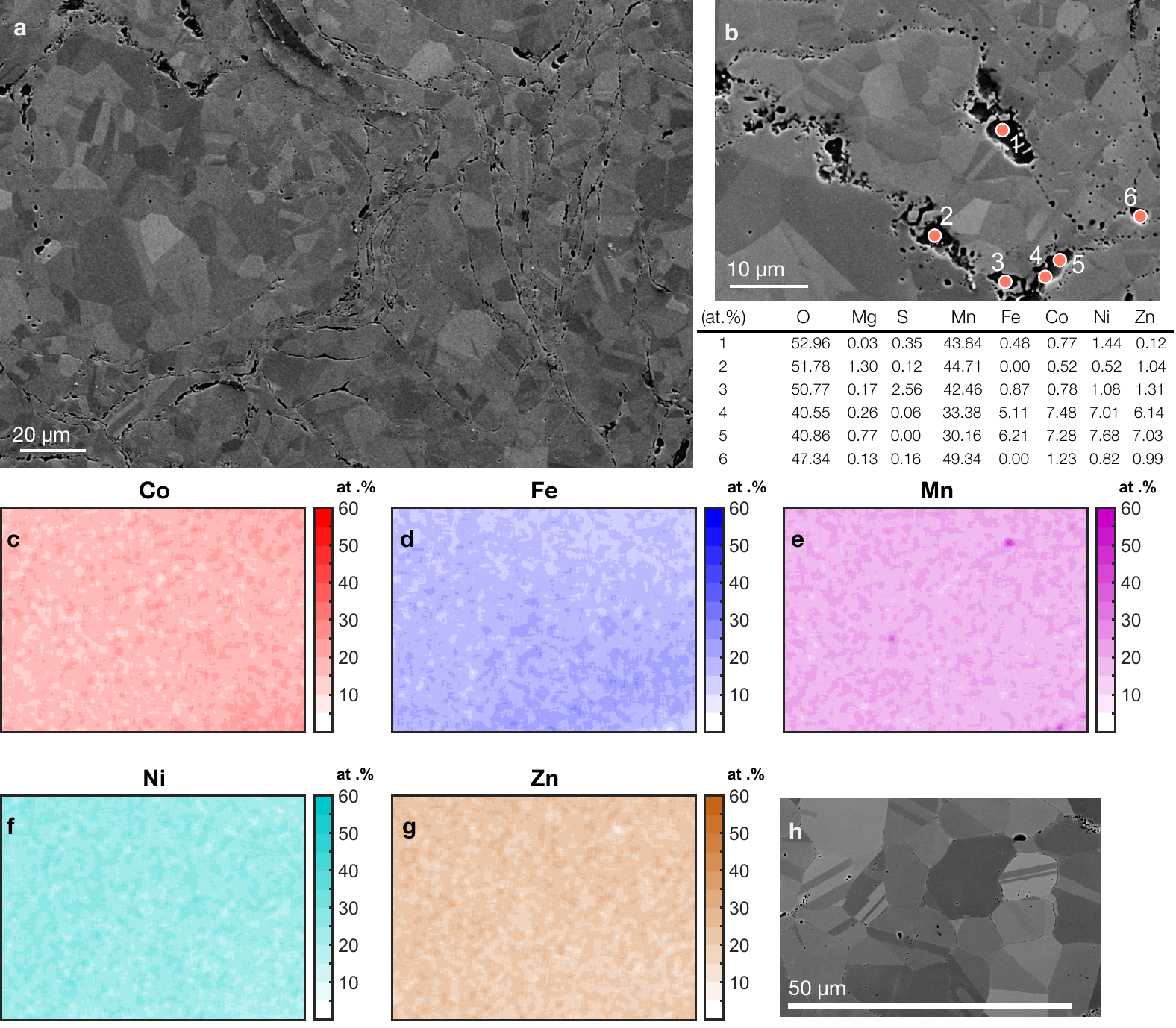}
\caption{\label{fig:supFCC} \textbf{a} Low magnification (SE) SEM micrograph of \ce{CoFeMnNiZn}. The single phase FCC grains are decorated by oxides. \textbf{b} EDX measurements for a selected number of oxide particles. \textbf{c-g} EDX mapping of the FCC matrix of region \textbf{h}.}
\end{figure*}

\clearpage

\bibliography{main}